\documentclass[twocolumn,showpacs,preprintnumbers,prb]{revtex4-1}
\usepackage{dcolumn}
\usepackage{graphicx}
\usepackage{amsmath, amssymb}
\usepackage{bm}
\usepackage{mathrsfs}

\newcommand{\tprimemax}{\ensuremath{t'_{\mbox{\scriptsize max}}}}
\newcommand{\tmax}{\ensuremath{t_{\mbox{\scriptsize max}}}}
\newcommand{\fnoise}{f_{\mbox{\scriptsize noise}}}
\newcommand{\Fnoise}{F_{\mbox{\scriptsize noise}}}
\newcommand{\tildeFnoise}{\tilde{F}_{\mbox{\scriptsize noise}}}
\newcommand{\calN}{{\cal N}}

\begin{document}

\title{A semi-classical approach to electron spin resonance
in quantum spin systems}
\author{Shunsuke C. Furuya and Masaki Oshikawa}
\affiliation{Institute for Solid State Physics, University of Tokyo,
Kashiwa 277-8581, Japan}
\author{Ian Affleck}
\affiliation{Department of Physics and Astronomy, University of British
Columbia, Vancouver, BC, Canada V6T 1Z1}

\begin{abstract}
We develop a semi-classical approximation to electron spin resonance in
quantum spin systems, based on the rotor or non-linear
sigma model.
The classical time evolution is studied using
molecular dynamics while random initial conditions are
sampled using classical Monte Carlo methods.
Although the approximation may be especially powerful in two dimensions,
we apply it here to one-dimensional systems of large spin at intermediate
temperatures, in the presence of staggered and uniform
magnetic fields.
We first test the validity of the semi-classical approximation
by comparing the magnetization to quantum Monte Carlo results
on $S=2$ chains.
Then we calculate the ESR spectrum,
finding broad coexisting paramagnetic and spin wave resonances.
\end{abstract}
\pacs{76.20.+q, 75.10.Pq, 03.65.Sq}
\date{\today}
\maketitle

\section{\label{intro}Introduction}

Electron spin resonance  probes the dynamics of interacting spin 
systems at zero wave-vector in a magnetic field. The 
intensity of adsorption of microwave radiation of frequency $\omega$ is
$\propto \omega G^R_{\alpha \alpha}(\omega )$
where $ G^R_{\alpha \alpha}(\omega )$ is the retarded Green's function 
of $S_T^\alpha \equiv \sum_{\bm r}S^\alpha_{\bm r}$, the $\alpha$ component of 
the total spin operator.
Since $[S_T^\alpha ,H_0]=0$ for an SU(2) invariant 
Hamiltonian, $H_0$, such as the Heisenberg model,
the ESR spectrum remains exactly the same as that for a single spin.
Thus ESR is a highly sensitive probe to 
small anisotropies in the Hamiltonian.
In order to discuss the effect of small anisotropies,
the calculation of ESR spectra has to be very precise.
An approximation, that may be reasonable to study
other properties such as inelastic neutron scattering spectrum,
might lead to an errorneous change of ESR,
even in the absense of anisotropy.
Then it could not be applied to ESR.
This is a challenging aspect of ESR theory,
especially in strongly correlated systems.

Older theories of ESR\cite{KuboTomita54,MoriKawasaki62}
are based on high temperature 
expansions at $T\gg J$, the exchange coupling, where each spin 
behaves approximately independently or else on spin-wave theory at $T<T_N$, 
the N\'eel temperature, typically of order $J$.  For the two dimensional Heisenberg model on the square lattice, $T_N=0$, 
leaving a large temperature range where neither of these approaches to 
ESR applies. In this ``renormalized classical region'' an approach to 
dynamics was developed by Chakravarty, Halperin and Nelson\cite{CHN}
(CHN) based 
on the two dimensional classical rotor model (CRM).
The CRM is defined with
two vector variables $\bm n_{i}$ and $\bm L_{i}$ at each site,
and the Hamiltonian
\begin{equation}
 \mathcal H = - b^{D-2}\rho_s\sum_{\langle i,j\rangle}
   \bm n_i \cdot
   \bm n_{j} + \frac {b^D}{2 \chi_{u\perp}}\sum_i \bm L_i^2,
   \label{rotor}
 \end{equation}
where
$D$ is the dimension of the lattice ($D=2$ in Ref.~\onlinecite{CHN})
and $b$ is the lattice constant.
$\bm n_i$ and $\bm L_j$ are subject to constraints
\begin{align}
 {\bm n_i}^2 & = 1, \\
 \bm n_i \cdot \bm L_i &= 0 .
\end{align}
Physically, $\bm L_i$ represents the angular momentum
of the rotor $\bm n_i$, and
the two parameters $\rho_s$ and $\chi_{u\perp}$
represent, respectively, the spin stiffness and the transverse component
of uniform susceptibility.

The dynamics of the system is defined by the
Hamiltonian~\eqref{rotor} together with the Poisson brackets
\begin{equation}
\begin{aligned}
 \{ L^\alpha_i, L^\beta_j \} &= \sum_\gamma
 \epsilon^{\alpha \beta \gamma} \delta_{ij} L^\gamma_i , \\
 \{ L^\alpha_i, n^\beta_j \} &= \sum_\gamma
 \epsilon^{\alpha \beta \gamma} \delta_{ij} n^\gamma_i , \\
 \{ n^\alpha_i, n^\beta_j \} &= 0,
\end{aligned}
\label{eq:Poisson}
\end{equation}
where $\alpha,\beta,\gamma = 1,2,3$ and $\epsilon^{\alpha \beta \gamma}$
is the Levi-Civita symbol.

The classical time evolution is calculated
using molecular dynamics simulations starting from random initial 
conditions which are generated with classical Monte Carlo methods. 
CHN and Ty\v{c}, Chakravarty and
Halperin\cite{TycHalperinChakravarty-PRL1989}
applied this method to calculate the neutron scattering cross-section 
at wave-vector near the antiferromagnetic point $(\pi /a,\pi /a)$,
where $a$ is the lattice constant of the spin system.
The purpose of this paper is to explore the applicability of this 
method to dynamics at zero wave-vector, which is relevant
to ESR. 

In fact, as shown in Refs.~\onlinecite{DamleSachdev98,BuragohainSachdev99},
this classical method can also be applied in dimension $D=1$, at least 
for sufficiently large spin magnitude, $S$. While N\'eel order doesn't 
occur even at $T=0$ for the one dimensional Heisenberg model there 
is another important characteristic low energy scale, $\Delta$ of order 
\begin{equation}
\Delta \approx Je^{-\pi S}.\label{Delta}
\end{equation}
For integer $S$ this is the energy of the gapped triplet magnons, predicted 
by Haldane and following from the behavior of the quantum O(3) 
non-linear sigma model (NLSM), basically the continuum limit of Eq. (1). 
At low temperatures, $T\ll \Delta$
ESR can be calculated\cite{Affleck90, HuangAffleck04} from transitions
between single magnon states (or, if appropriate symmetry breaking 
is included in $H$, from magnon production processes). Magnon-magnon 
interactions become unimportant in this temperature range because 
the magnons are dilute, with density $\propto e^{-\Delta /T}$.  
Instead, at higher temperature, the magnons are dense,
and the NLSM behaves rather classically.
The rotor model is nothing but a lattice version of the NLSM;
thus CRM describes the classical dynamics of the NLSM.

The present approach to ESR based on the 
CRM can be used in the intermediate temperature range
\begin{equation}
 \Delta \ll T \ll J S^2,
\label{Trange1}
\end{equation}
where the magnons are dense
and the CRM is still valid as an effective theory.
Unfortunately, this temperature range may not exist for the
spin $S=1$ chain, for which the Haldane gap is known to be
$\Delta \sim 0.41 J$.
However, the gap becomes smaller for higher $S$ as in eq.~\eqref{Delta}
and the temperature range~\eqref{Trange1} 
becomes well-defined for higher spins, perhaps starting at $S=2$,
for which the Haldane gap is already as small\cite{Todo01} as
$\Delta \sim 0.089 J$.

While the classical method has a wider range of $T$ and $S$ over which
it is applicable in $D=2$, we focus on the $D=1$ case in this paper, due
to the computational cost of the classical technique, which appears to
be quite severe for ESR studies.  We hope to return to the $D=2$ case in
the future.
Therefore, 
in this paper, we discuss ESR in 1D systems in
the previously unexplored temperature regime~\eqref{Trange1}.
We will discuss conditions for the present approach
to be applicable, in more detail, in
Sec.~\ref{sec:conditions}.

Following Refs.~\onlinecite{TycHalperinChakravarty-PRL1989,BuragohainSachdev99},
initial states are generated by classical Monte
Carlo simulation of the O(3) CRM at given temperature.
The real-time correlation function is then obtained by solving
the classical equation of motion for the O(3) CRM
with the initial condition.
ESR spectrum can be obtained from
Fourier transform of the real-time correlation function.

Among various possible anisotropies, in this paper
we discuss the staggered field, which is known to have
most interesting effects on ESR spectra
in $S=1/2$ and $S=1$ cases.
The effective Hamiltonian including the staggered field
is given by
\begin{equation}
 \mathcal H = \sum_j J \bm S_j \cdot \bm S_{j+1} - HS_j^z -(-1)^j h
  S_j^x,
  \label{QSH}
\end{equation}
where $H$ and $h$ is the uniform field and the transverse staggered
field respectively.
Throughout this paper, we set $g \mu_B = \hbar = k_B =1$ for
simplicity.

While the staggered field might seem unphysical,
it often does appear effectively in actual
quantum antiferromagnetic chains.
When an external magnetic field is applied to
a material with a staggered crystal structure 
along the chain, the staggered field is effectively
generated through a staggered $g$ tensor\cite{SakaiShiba94,Mitra94}
and also through
a staggered Dzyaloshinskii-Moriya interaction.\cite{Oshikawa97}
Examples of such materials include the typical $S=1$ Haldane chain
material $\mathrm{Ni(C_2H_8N_2)_2NO_2ClO_4}$ (NENP)\cite{Meyer82},
and the $S=1/2$ Heisenberg antiferromagnetic chain
Cu benzoate\cite{Date-Cubenz}.

For ESR in the $S=1/2$ antiferromagnetic chain at low temperatures,
field theory approach leads to diverging linewidth of the
paramagnetic peak at lower temperatures, and an appearance
of a new peak when the temperature is lowered
further down to zero\cite{Oshikawa99,OshikawaAffleck02}.
For ESR in the $S=1$ Haldane chain at low temperatures,
the violation of selection rule by the staggered field
leads to the appearance of a new peak at the frequency
equal to the Haldane gap,
corresponding to creation of
a single magnon\cite{SakaiShiba94,Mitra94}.
These approaches are only justified in a ``quantum'' regime
and no longer valid for a Haldane chain at the intermediate
temperatures~\eqref{Trange1}, which is the focus
of the present paper.

The $S=2$ Heisenberg antiferromagnetic chain, for
which the present approach would be relevant,
is not just a theoretical toy model.
Granroth {\it et al.} \cite{Granroth96} reported
an experimental evidence of Haldane gap in 
an $S=2$ antiferromagnetic Heisenberg chain compound
$\mathrm{MnCl_3(bpy)}$ (bpy $=$ bipyridine).
$\mathrm{MnCl_3(bpy)}$ is similar to NENP.
It has a quite small single ion anisotropy $D/J \le 0.04 \pm 0.02$
and has a staggered crystal structure.
The staggered crystal structure would cause the staggered
$g$ tensor, which produces an effective staggered field when
an external magnetic field is applied.
Thus it would be interesting to measure ESR in
$\mathrm{MnCl_3(bpy)}$ and compare to the
present theory.

Our results may be applied to a wider range of one-dimensional
systems such as quantum spin ladders as well,
since they can also be described
by the same CRM\cite{BuragohainSachdev99}.
For spin ladders, there are various possible generalizations
of the staggered field.
The results of the present paper can be directly
applied when the staggered field
is unfrustrated\cite{Sato-coupled-staggered,Zhao-ladder-staggered},
for example as in
\begin{align}
 \mathcal H = &  \sum_{\mu=1}^n \sum_{j}
\left[
J \bm S_{\mu,j} \cdot \bm S_{\mu, j+1}
             - HS_{\mu,j}^z -(-1)^{\mu+j} h  S_{\mu,j}^x,
\right]
\notag \\
&
+ \sum_{\mu=1}^{n-1} \sum_j J_{\perp} \bm S_{\mu,j} \cdot \bm S_{\mu+1,j},
\label{eq:H_ladder}
\end{align}
where $n$ is the number of legs and $\mu$ is the leg index.
In this case, the staggered field can be
handled in the same way as in the chain~\eqref{QSH}.

In fact, even for gapless one-dimensional systems
such as half-integer spin Heisenberg antiferromagnetic chains,
$\Delta$ defined in Eq.~(\ref{Delta}) is a characteristic energy scale.
At energy
scales of order $\Delta$, the system renormalizes from the weak coupling
regime, where classical methods can be used, to the non-trivial critical
point induced by the topological term in the effective Lagrangian.  At
temperatures small compared to $\Delta$ the low temperature theory for
the $S=1/2$ Heisenberg model\cite{Oshikawa99,OshikawaAffleck02} can be
used.  (For the $S=1/2$ case, this theory is valid at any temperature
$T\ll J$.)
On the other hand, in the temperature range~\eqref{Trange1},
the present approach based on the CRM is valid even
in gapless systems.

Physically, ESR in the presence of the staggered
field provides an interesting case of crossover of dynamics
between two different regimes.
When the effect of the staggered field is weak,
the ESR spectrum is dominated by the
paramagnetic resonance at $\omega \sim H$, and
the staggered field causes its broadening and shift.
However, when the effect of the staggered field is strong,
the system is ordered along the staggered field.
The ESR spectrum is dominated by the spin-wave type fluctuation
around the ordered state.
This is similar to the case in which the system has a
N\'{e}el order spontaneously,
but is different in that the ``order'' is imposed externally
by the staggered field.
Nevertheless,
theory of antiferromagnetic resonance\cite{Nagamiya-AFMR,Keffer-Kittel-AFMR}
developed
for the spontaneously ordered state can be modified
and applied to the present case, as
discussed for example in Ref.~\onlinecite{SakaiShiba94}.
It describes the limit of the strong staggered field,
where the imposed N\'{e}el order is perfect.

In the case of a spontaneous ordering, a phase transition
separates the ordered and disordered phases.
In the present case, there is no phase transition but
only a smooth crossover between the two regimes.
The description of the crossover is generally much more
difficult than that of the limiting cases.

For the $S=1/2$ chain, the field theory approach successfully
gives the broadening and shift of the paramagnetic resonance,
when the staggered field is small (or the temperature
is sufficiently high).
It also explains the ESR spectrum in the
low temperature limit, when the system is ordered along
the staggered field.
The antiferromagnetic resonance in this case is renormalized
due to strong quantum fluctuations in the $S=1/2$ chain.
This can be well described in terms of elementary excitations
of quantum sine-Gordon field theory.
However, despite the integrability of the quantum sine-Gordon
field theory, dynamical susceptibility at finite temperature
has not been obtained exactly.
Thus the theoretical description of the crossover between
two regimes still remains unsolved, although a numerical result
based on exact diagonalization was reported\cite{Iitaka-ESR}.

In other systems, the crossover is even less understood.
The present approach can, where it is valid,
numerically describe the nontrivial crossover of the ESR
spectrum between two regimes as we will demonstrate.

This paper is organized as follows.
In Sec. \ref{NLSM}, we introduce the quantum O(3) NLSM
as an effective low-energy theory for the Hamiltonian~\eqref{QSH},
and review its renormalization.
ESR spectrum is also formulated in terms of the O(3) NLSM.
We will then discuss the classical approximation
of the O(3) NLSM, including the range of validity
of the approximation, in Sec.~\ref{CA}.
The CRM is introduced as a lattice version, suitable
for numerical calculation, of the classical limit of the O(3) NLSM.
In Sec.~\ref{ESR} we will obtain ESR spectra
by numerically solving the equation of motion
for the CRM.
We will develop a theory of antiferromagnetic
spin-wave resonance in Sec.~\ref{SW},
which can be identified with the new resonance
at higher frequency found in the numerical results.
Finally in Sec. \ref{summary} we summarize the paper.

\section{\label{NLSM}O(3) nonlinear sigma model}

\subsection{Definition of the model}
\label{sec:NLSMdef}

Let us introduce the
O(3) NLSM as an effective field theory of the
antiferromagnetic Heisenberg chain,
and summarize its properties.
Although we will eventually treat the system classically,
first we need to clarify the effects of quantum fluctuation
in order to determine the appropriate parameters
in the effective classical model.

The O(3) NLSM is defined in terms of
two fields $\bm n(x)$ and $\bm L(x)$ which are related to the
original spin $\bm S_j$ through
\begin{equation}
 \bm S_j = (-1)^j S \bm n(x) \sqrt{1- \biggl( \frac{a\bm
  L(x)}{S}\biggr)^2} + a\bm L(x).
  \label{S2Ln}
\end{equation}
Here $x=ja$, and $a$ is a lattice spacing.
Hereafter we set the lattice spacing to  $a =1 $.
$\bm n(x)$ and $\bm L(x)$ satisfy constraints
\begin{align}
\bm{n}^2(x) & = 1,
\label{n2constraint}
 \\
\bm{n}(x) \cdot \bm{L}(x) &= 0,
\end{align}
which are necessary to keep the
constraint $\vec{S}^2 = S(S+1)$.
Then $\bm n(x)$ and $\bm L(x)$ satisfy the following commutation
relations
\begin{align}
 [L^\alpha (x), L^\beta (y)]&=i\epsilon^{\alpha \beta \gamma}
 L^\gamma(x)  \delta(x-y), \label{LL=L} \\
 [L^\alpha (x), n^\beta (y)]&=i\epsilon^{\alpha \beta \gamma}
 n^\gamma(x)  \delta (x-y), \label{Ln=n} \\
 [n^\alpha (x), n^\beta (y)]&= 0. \label{nn=0}
\end{align}
O(3) NLSM is derived after substituting \eqref{S2Ln} to \eqref{QSH} and
taking the continuum limit.
The Hamiltonian density of O(3) NLSM is given by
\begin{equation}
 \mathscr H = \dfrac c{2 g} (\partial_x
  \bm n)^2 + \dfrac{c g}2 \bm L^2 
  - HL^z   -\dfrac{\Delta_h^2}{gc} n^x,
\label{NLSMH}
\end{equation}
where $c=2JS$ is the spin-wave velocity,
$g = 2/S$ is the coupling constant,
and $\Delta_h$ is defined as
\begin{equation}
 \Delta_h = \sqrt{4JSh},
 \label{eq.Deltah}
\end{equation}
Later in this Section,
we will show that $\Delta_h$ is
the staggered-field-induced gap.
The coupling constant $g$ is actually subject to
renormalization owing to quantum fluctuations,
which we will discuss in the next subsection.
$g=2/S$ should be understood as the bare coupling constant.

Here we introduce useful normalization which measure the energy scale 
in units of $c=2JS$.
\begin{subequations} 
 \begin{align}
  \mathcal H' & = \frac{\mathcal H}{2JS}
  \label{eq.Ham'} \\
  T' &= \frac T{2JS}
  \label{eq.T'} \\
  H' &= \frac H{2JS}
  \label{eq.H'}\\
  h' &= \frac h{2JS}
  \label{eq.h'} \\
  \Delta'_h &= \frac{\Delta_h}{2JS} = \sqrt{2h'}
  \label{eq.Delta'_h} \\
  t' &= 2JS \, t
  \label{eq.t'} \\
  \omega' &= \frac{\omega}{2JS}
  \label{eq.omega'}
\end{align}
\end{subequations}
We specify the normalized, dimensionless parameters
by adding the prime symbol as in the above.
In this notation, the Hamiltonian \eqref{NLSMH} is rewritten as
\begin{equation}
 \mathscr H' = \frac 1{2g} (\partial_x \bm n)^2
  + \frac g2 \bm L^2 - H'L^z - \frac{\Delta'^2_h}{g} n^x.
  \label{NLSMH'}
\end{equation}

Because Eq. \eqref{NLSMH} (or Eq. \eqref{NLSMH'}) is harmonic in
$\bm L(x)$, $\bm L(x) - \bm H/gc = \bm n \times (\partial_t \bm n + \bm
H \times \bm n)/gc = \bm n \times (\partial_{t'} \bm n + \bm H' \times
\bm n)/g$ can be integrated out.
Then the Lagrangian is given by
\begin{align}
 \mathscr L' &= \partial_{t'} \bm n \cdot (\partial_{t'} \bm n + \bm H'
 \times \bm n) - \mathscr H' \notag \\
 &= \frac 1{2g}(\partial_{t'} \bm n + \bm H' \times \bm n)^2
 - \frac 1{2g} (\partial_x \bm n)^2 + \frac{\Delta'^2_h}{2g} n^x,
 \label{NLSML}
\end{align}
with the constraint~\eqref{n2constraint}.
In this expression, it is clear that the coupling constant $g$
represents the degree of quantum fluctuation,
since it plays the role of Planck's constant.

Although there is no energy scale in the Lagrangian density,
O(3) NLSM at $H=h=0$ has an excitation gap, which corresponds
to the Haldane gap $\Delta$.
In field theory language, the mass gap is dynamically
generated.
The low-energy phenomena of the O(3) NLSM can be
effectively described by the field theory of triplet
bosons with the mass $\Delta$:
\begin{equation}
 \mathscr L
  =  \frac c{2g} (\partial_\mu \bm n)^2
  - \frac{1}{2gc}\Delta^2 \bm{n}^2(x),
  \label{GLLag}
\end{equation}
{\em without} the constraint~\eqref{n2constraint}.
Hereafter $(\partial_\mu \bm n)^2$ means
$(\partial_\mu \bm n)^2 = (1/c^2)(\partial_{t} \bm n)^2
-(\partial_x \bm n)^2 = (\partial_{t'} \bm n)^2 - (\partial_x \bm n)^2$.

When the applied field $H$ becomes larger than $\Delta$ 
while the staggered field $h=0$,
the gap is closed.
On the other hand, when staggered field $h$ is large,
the excitation gap is still open even if $H>\Delta$
because the Lagrangian \eqref{NLSML}
is approximated as follows.
\begin{equation}
  \mathscr L' = \frac 1{2g} (\partial_\mu \bm m)^2
 -\frac{\Delta'^2_y}{2g}(m^y)^2
 - \frac{\Delta'^2_z}{2g} (m^z)^2,
  \label{CNLSML}
\end{equation}
where $\Delta'_{\alpha} = \sqrt{H'^2 \delta_{\alpha, z} + {\Delta'_h}^2}$.
The field
$\bm m = (0, n^y, n^z)$ contains the components of $\bm n$ transverse 
to the  staggered field direction.
We assumed $n^x(t,x) \approx 1$
because of large $h$.
$\bm m$ represents the elementary excitation in the large $h$ regime
and has a dispersion
${E'_k}^{(\alpha)} = \sqrt{k^2 + {\Delta'_{\alpha}}^2}$ where $k$ is the
wave number and $\alpha = y, z$.
This dispersion relation indicates that the gap for the $z$ component
is different from the gap for the $y$ component:
\begin{align}
 \Delta'_z &= \sqrt{{H'}^2 + {\Delta'_h}^2},
 \label{eq.Deltaz} \\
 \Delta'_y &= \Delta'_h, 
 \label{eq.Deltay}
\end{align}
where $\Delta'_h$ is the staggered-field-induced gap~\eqref{eq.Delta'_h}.
The gap \eqref{eq.Deltaz} is
derived within the
classical approximation ignoring quantum fluctuation.
On the other hand, in the absence of the staggered field,
the quantum fluctuation leads to the Haldane gap $\Delta$.
Roughly speaking, the excitation gap of the system is given as
\begin{equation}
  {\tilde \Delta}' \sim \mathrm{max}(\Delta', \Delta'_h)
   \label{ActualGap}
\end{equation}

In ESR measurements, we apply an oscillating field with frequency
$\omega$ and polarization $\alpha$.
In this paper we consider Faraday configuration where the polarization
is perpendicular to the direction of the uniform field.
ESR absorption intensity
\begin{equation}
I(\omega) \propto \omega \chi''_{\alpha\alpha}(\omega),
\label{eq:def_Intensity}
\end{equation}
is written in terms of the dynamical susceptibility
$\chi_{\alpha\alpha}(\omega)$.
According to the linear response theory, the imaginary part of
$\chi_{\alpha\alpha}(\omega)$ is related to Fourier
component of the retarded Green function
$G^R_{\alpha\alpha}(t)=-i\theta(t) \langle
[L^\alpha_{\mathrm T}(t), L^\alpha_{\mathrm T}(0)] \rangle$, namely
$\chi''_{\alpha \alpha}(\omega) = -\operatorname{Im}G^R_{\alpha\alpha}(\omega)$.
The retarded Green function
$G^R_{\alpha\alpha}(\omega)$ is defined as
\begin{equation}
 G^R_{\alpha\alpha}(\omega) = -i\int_0^\infty dt \,
  e^{i\omega t} \langle[ L^\alpha_{\mathrm T}(t),
  L^\alpha_{\mathrm T}(0)] \rangle.
  \label{GRaa}
\end{equation}
$L_{\mathrm T}^\alpha = \int dx \, L^\alpha(x)$ is the total
magnetization.
In general, the ESR spectrum thus depends on the polarization $\alpha$.

In Ref.~\onlinecite{OshikawaAffleck02}, it was shown that
for the system~\eqref{QSH}
with the staggered field in $x$-direction,
the polarization dependence can be exactly determined by
the equation of motion:
\begin{equation}
 \chi''_{\alpha\alpha}(\omega) = \frac{H^2\cos^2\Phi+ \omega^2
  \sin^2\Phi}{\omega^2}\chi''_{yy}(\omega),
\end{equation}
where $\Phi$ is the angle between the polarization $\alpha$
and $x$-axis.

Furthermore, ESR spectrum for circular polarization can be
related to
\begin{equation}
 \chi''_{+-}(\omega) =\operatorname{Re} \int_0^\infty dt \,
  e^{i\omega t} \langle [L^+_{\mathrm T}(t), L^-_{\mathrm T}(0)]
  \rangle,
  \label{eq.chi+-_comm}
\end{equation}
where $L^\pm = L^x\pm i L^y$.
The discussion in Ref.~\onlinecite{OshikawaAffleck02} can be
extended to circular polarization to obtain
\begin{equation}
 \chi''_{+-}(\omega) = \biggl(1 + \frac{H}{\omega}\biggr)^2 \chi''_{yy}(\omega) .
\end{equation}
That is,
ESR spectra for different polarization in Faraday configuration
are related to each other, and thus
it is sufficient to compute the spectrum for one particular case.

In this paper, we consider
the circular polarization case, eq.~\eqref{eq.chi+-_comm}.
A short calculation yields
\begin{equation}
 \chi''_{+-}(\omega) = (1-e^{-\omega/T}) \operatorname{Re}
  \int_0^\infty dt \, e^{i\omega t} \langle
  L^+_{\mathrm T}(t) L^-_{\mathrm T}(0) \rangle,
  \label{chi+-}
\end{equation}
which is convenient for our purpose since eq.~\eqref{chi+-}
is well-defined even if $L^\alpha(t)$ is a classical
vector.
In this paper, we numerically solve
the classical equation of motion to obtain
the dynamical correlation function
\begin{equation}
 \langle L^+_{\mathrm T}(t) L^-_{\mathrm T}(0) \rangle .
\label{eq:LLcorr}
\end{equation}
This gives the desired ESR spectrum by eq.~\eqref{chi+-}.
Our numerical approach to this problem will be discussed
in Sec. \ref{CA} and Appendix~\ref{sec.numerics}.

\subsection{\label{RG}Renormalization group transformation}

Let us consider the renormalization group transformation
of general O($N$) NLSM whose Lagrangian is
given by
\begin{align}
 \mathscr L' &= \frac 1{2g} \bigl(\partial_{t'} \bm n + \bm H' \times \bm n
  \bigr)^2 - \frac 1{2g} (\partial_x \bm n)^2 +
  \frac{2h'}{g}n^x.
\end{align}
This bare theory is defined at the energy scale $J$.
we renormalize this theory down to an energy scale $E<J$.

When $H=h=0$, 
the effective coupling $g_R(E)$ at energy scale $E$
satisfies \cite{BZJ76}
\begin{equation}
 \frac{1}{g} - \frac 1{g_R(E)} = \frac{N-2}{2\pi} \ln \biggl(
  \frac EJ \biggr).
 \label{RGeqforg}
\end{equation}
As the bare coupling satisfies $1/g = (N-2)\ln (J/\Delta)/2\pi$,
we can simplify the renormalized coupling constant
\begin{equation}
 \frac 1{g_R(E)} = \frac {N-2}{2\pi} \ln \frac E\Delta.
\end{equation}
For the case of our interest, $N=3$, we find
\begin{equation}
  g_R(E) = \frac{2\pi}{\ln (E/\Delta)} .
  \label{renormalizedg}
\end{equation}

Next we consider the renormalization of the staggered field.
We ignore the influence of $h'$ on the renormalization of $g$,
but consider the renormalization of $h'$ produced by $g$, at
lowest order.
This is determined by the anomalous dimension of the field $\bm n$
\cite{BZJ76, Affleck89}.
\begin{equation}
 \frac{\partial h'_R}{\partial \ln E} = -\gamma\big(g_R(E)\big) h'_R
  \label{dhdE}
\end{equation}
To one loop order, the anomalous dimension $\gamma(g)$ is
\begin{equation}
  \gamma(g) = - \frac{N-3}{4\pi} g
   \label{AnomalousDimension}
\end{equation}
Using Eq. \eqref{RGeqforg} for $g_R(E)$, Eq. \eqref{dhdE}
becomes:
\begin{equation}
 \frac{d\ln h'_R}{d\ln E} = -\frac{N-3}{2(N-2)} \frac 1{\ln (E/\Delta)}
  \label{dlnhdlnE}
\end{equation}
This gives the renormalized staggered field
\begin{equation}
 h'_R(E) = h'
  \biggl[ \frac{\ln (E/\Delta)}{\ln (J/\Delta)}\biggr]
  ^{(N-3)/2(N-2)}.
  \label{hR(T)}
\end{equation}
Namely, for the case of our interest $N=3$,
the renormalization of the staggered field $h'$ is absent.
The renormalized Lagrangian is given by
just replacing $g$ by $g_R(E)$ in eq.~\eqref{NLSML}.
Because $h'_R(E) = h'$ is kept unchanged
(within the leading order considered in this paper)
for $N=3$, there is no logarithmic correction in the
staggered-field-induced gap \eqref{eq.Deltah}

\section{\label{CA} Classical Approximation}

\subsection{Conditions for the classical approximation}
\label{sec:conditions}

In the Introduction,
we have briefly discussed the condition~\eqref{Trange1}
for the classical approximation being justified.
Here we discuss the condition in some more detail,
with several additional conditions.

The latter condition in eq.~\eqref{Trange1},
\begin{equation}
  T \ll JS^2,
\label{eq:T_ll_JS2}  
\end{equation}
is required for the
validity of the effective theory in the continuum.
It is equivalent to the requirement that the correlation
length of the antiferromagnetic order parameter
is much longer than the lattice spacing.
In fact, as pointed out in Ref.~\onlinecite{BuragohainSachdev99},
two different temperature regimes can be distinguished
for large $S$.
\begin{equation}
  T < T_{\mbox{\scriptsize max}}^{(1)} \sim 2 JS ,
\label{eq:T_Tmax1}
\end{equation}
and
\begin{equation}
 T_{\mbox{\scriptsize max}}^{(1)} \sim 2 JS < T
  < T_{\mbox{\scriptsize max}}^{(2)} \sim JS^2 .
\label{eq:T_Tmax2}
\end{equation}
In the regime~\eqref{eq:T_Tmax1}, the O(3) NLSM description
is valid.
On the other hand, when eq.~\eqref{eq:T_Tmax2} holds,
the quantum spin chain may be directly
approximated as a classical spin chain and
then the continuum description is applied.
The resulting dynamics is equivalent to that
of the classical O(3) NLSM, although
the effective parameters are estimated in a
different manner.
For simplicity, throughout the rest of this paper,
we assume the regime~\eqref{eq:T_Tmax1}.
Although ESR in the higher temperature range
$ T_{\mbox{\scriptsize max}}^{(2)} \sim JS^2 \lesssim T$
is also an interesting nontrivial problem, it is out of scope
of the present paper.

ESR is studied under an applied magnetic field $H$.
If the ground state is fully polarized along $H$,
O(3) NLSM is not appropriate as an effective theory.
Thus, our approach also requires that
the uniform field should be much weaker than
its saturation field, which is order of $JS$:
\begin{equation}
 H \ll JS  .
\label{eq.Hrange}
\end{equation}

The first inequality in eq.~\eqref{Trange1} is
necessary for the classical approximation to hold.
Namely, that the temperature higher than the gap,
implies high density of thermally excited magnons.
This leads to a breakdown of the quantum mechanical
picture\cite{Affleck90} of ESR based on transitions of
independent magnons.
In fact, as we have discussed in Sec.~\ref{sec:NLSMdef},
the staggered field induces a gap as in eq.~\eqref{eq.Deltah}.
Since actual gap of the system is given as
eq.~\eqref{ActualGap},
the first inequality in eq.~\eqref{Trange1}
should be replaced by
\begin{equation}
 {\tilde{\Delta}} = \mathrm{max}(\Delta,\Delta_h) \ll T. 
\end{equation}

Namely, the temperature must be higher than not only the Haldane
gap $\Delta$, but also the staggered-field induced gap $\Delta_h$.
Together with eqs.~\eqref{eq:T_ll_JS2} and~\eqref{eq.Deltah},
this requires
\begin{equation}
  h \ll \frac{T^2}{4JS}  \ll \frac{JS^3}{4} .
\label{eq:h_for_classical}
\end{equation}
Later, we will discuss an antiferromagnetic spin-wave theory of ESR.
It is justified when
\begin{equation}
  h \gg \frac{\pi^2 T^2}{4JS \bigl(\ln(T/\Delta)\bigr)^2},
\label{eq:h_for_SWT}
\end{equation}
where we have assumed eq.~\eqref{eq:H_ll_T}.
For a larger spin $S$, the Haldane gap $\Delta$ is expected
to become exponentially small as in eq.~\eqref{Delta}.
Thus the range of $h$ satisfying both eqs.~\eqref{eq:h_for_classical}
and~\eqref{eq:h_for_SWT} becomes wide for large $S$.
In fact, we will demonstrate that the spin-wave theory
prediction agrees very well for a $S=10$ chain with
staggered field, in a range of parameters.
On the other hand, for $S=2$, when the classical approximation
is valid, the spin-wave theory is not quite justified.
Correspondingly, a broad peak is observed instead of a sharp
resonance. Nevertheless, the broad peak may be understood as
a remnant of the spin-wave resonance.

For discussion of ESR based on the classical dynamics
of the NLSM,
the frequency $\omega$ of the applied oscillating field
also should be much lower than the temperature.
\begin{equation}
 \omega \ll T
  \label{eq.w_ll_T}
\end{equation}
This also implies
\begin{equation}
 H \ll T ,
\label{eq:H_ll_T}
\end{equation}
because ESR absorption
usually occurs for $\omega \geq H$.

\subsection{Effective Hamiltonian}

In the Introduction,
we argued that the system should be described by
the classical O(3) NLSM in the
intermediate temperature range~\eqref{Trange1}.

Even in the temperature range~\eqref{Trange1},
we cannot simply ignore quantum fluctuations at energy scale
above the temperature $T$. 
Their effects are  taken into account by
the renormalization group.
The effective classical Hamiltonian for the O(3) NLSM
may be obtained by using the renormalized parameters,
eqs.~\eqref{renormalizedg} and~\eqref{hR(T)},
and setting the energy scale to the temperature: $E=T$.

On the other hand, we do not consider the renormalization of
$\bm H \cdot \bm L$ term since $\int dx \, \bm L$ is a conserved
quantity when $h=0$.
Therefore any renormalization that does occur should vanish at $h=0$
and would be negligible at small $h$.
The renormalized Hamiltonian density is thus given as
\begin{align}
 \mathscr H_{\mathrm{cl}}& = \frac c{2g_R} (\partial_x \bm n)^2
 + \frac{cg_R}2 \bm L^2
 - HL^z - \frac{\Delta_h^2}{2cg_R} n^x,
 \label{eq.CNLSMH}
\end{align}
where $g_R = g_R(T)$.
In terms of dimensionless parameters, it reads
\begin{equation}
 \mathscr H'_{\mathrm{cl}} = \frac 1{2g_R} (\partial_x \bm n)^2
  + \frac{g_R}2 \bm L^2 - H'L^z - \frac{2h'}{g_R}n^x.
  \label{eq.CNLSMH'}
\end{equation}

As we will discuss below,
the coefficient of the $(\partial_x \bm n)^2$ term and
the $\bm L^2$ term in
the classical Hamiltonian \eqref{eq.CNLSMH} can be
identified respectively with $\xi T/2$ and $1/2\chi_{u\perp}$,
where
$\xi$ is the classical correlation length of $\bm n$ 
and 
$\chi_{u \perp} = (3/2) \chi_u$ is proportional to the
zero-field uniform susceptibility $\chi_u$,
both {\em at zero fields} $H=h=0$.
Namely,
\begin{align}
 \xi &= \frac{c}{2\pi T} \ln  \frac T\Delta
 \label{xi} \\
 \chi_{u\perp} &= \frac 1{2\pi c} \ln \frac T\Delta
 \label{chi}
\end{align}
Thus, the effective Hamiltonian density~\eqref{eq.CNLSMH}
can be also written as
\begin{equation}
  \mathscr H_{\mathrm{cl}} =
 \frac{\xi T}2 (\partial_x \bm n)^2 + \frac 1{2\chi_{u \perp}}
 \bm L^2(x)
 - HL^z(x) -\chi_{u\perp} \Delta_h^2 n^x,
 \label{CNLSMH_v2}
\end{equation}
which is the form used in Ref.~\onlinecite{BuragohainSachdev99}.

In numerical calculations we discretize the effective field theory
\eqref{eq.CNLSMH'} and consider the Hamiltonian,
\begin{equation}
 \mathcal H'_{\mathrm{cl}} = \sum_j b\biggl[ - \frac 1{b^2 g_R} \bm n_j
  \cdot \bm n_{j+1} + \frac {g_R}2 \bm L_j^2
  -H'L^z_j - \frac{2h'}{g_R} \Delta_h^2 n^x\biggr],
  \label{CNLSMH_lattice}
\end{equation}
on a lattice with the lattice spacing $b$.
This is nothing but the CRM~\eqref{rotor} in dimension $D=1$.
We note that it is not necessary to take $b$ equal to
the lattice spacing $a=1$ of the original spin chain;
the CRM may be regarded as a lattice regularization
of the classical, continuum O(3) NLSM. 
Usually $b \geq a$ is taken, because
the eq.~\eqref{CNLSMH_lattice} is introduced to describe
long-distance assymptotic behavior of the spin system.
The system size $L= \calN b$ is proportional to the number of rotors
$\calN$.
We use $\calN=16$ for our numerical calculations.
As we will see in FIG.~\ref{fig.magQvsC},
$\calN=16$ is large enough to reproduce consistent values of the
magnetization density $M/L$ with quantum Monte Carlo calculations
and low-field expansion.
This is because the correlation length of $\bm n_j$ is 
much shorter than $L$ due to the relatively high temperature
$T \sim J$.

Let us now demonstrate that the above identifications are valid
within the classical theory.
To do so, we assume the Hamiltonian density in the form
of eq.~\eqref{CNLSMH_v2}, and then show that
$\xi$ and $\chi_u$ are indeed the correlation length
and the uniform susceptibility.
To calculate the correlation function of $\mathbf{n}$, 
it is convenient to integrate out $\bm L$ and
obtain the Hamiltonian in terms of $\mathbf{n}$.
For $H=h=\Delta_h=0$, it reads
\begin{equation}
 \mathcal H_{\mathrm{eff}} = \frac{T\xi}2 \int dx \, \biggl( \frac{d\bm
  n}{dx}\biggr)^2 \approx - \frac{T\xi}{b} \sum_j \bm n_j
  \cdot \bm n_{j+1}.
  \label{CHeisenbergH-zero_n}
\end{equation}
The last expression is the NLSM Hamiltonian on a discretized
one-dimensional lattice with the lattice constant $a$.
In fact, it is equivalent to
the classical Heisenberg chain.
The equilibrium statistical properties of the classical
Heisenberg chain are studied by Fisher~\cite{Fisher-classical}.
The correlation function
$\langle n^a(x)n^b(0) \rangle$ of the
classical Hamiltonian \eqref{CHeisenbergH-zero_n}
is  obtained exactly by using transfer matrix method,
$\langle n^a(x) n^b(0) \rangle = (1/3) \delta^{ab}
\exp(-|x|/\xi)$,
which indicates that Eq.~\eqref{xi}
is indeed the correlation length of the order
parameter $\bm n$.

Next we introduce an infinitesimal uniform field $\bm H$.
The zero-field uniform susceptibility $\chi_u$ is
defined as $\chi_u = \lim_{H \to 0} dM/dH$ where
$M = \int dx \langle L^z \rangle$ is the total
magnetization.
We calculate the uniform susceptibility $\chi_u$
of the classical NLSM~\eqref{CNLSMH_v2}, with
$h=\Delta_h=0$.
After integrating out $\bm L$, the Hamiltonian becomes
\begin{equation}
 \mathcal H = \int dx \biggl[ \frac{T\xi}2 \biggl(
  \frac{d\bm n}{dx}\biggr)^2 - \frac{\chi_{u\perp}H^2}2
  (\bm n^\perp)^2 \biggr].
  \label{CHeisenbergH_n}
\end{equation}
Here we defined $\bm n^\perp = (n^x, n^y, 0)$.
The uniform susceptibility is thus obtained as
\begin{equation}
 \chi_u = \lim_{H \to 0} \chi_{u\perp} \int dx \,
  \langle (\bm n^\perp)^2 \rangle
  = \frac 23 \chi_{u\perp}, 
  \label{chi_uperp2chi_u}
\end{equation}
where $\langle \rangle$ is the thermodynamic
expectation value at $H=0$.
$(\mathbf{n}^\perp)^2 = 2/3$ follows from
the isotropy of the Hamiltonian and
the constraint~\eqref{n2constraint}.
Thus, \eqref{chi} is proved to be a
transverse component of the uniform susceptibility $\chi_u$.

\subsection{\label{magnetization}Magnetization}

Our primary interest in this paper is in the dynamics of the system.
However, before going into the dynamics,
it would be important to establish the validity of the
present approach by considering static properties.
This was done earlier in
Ref.~\onlinecite{YJKim-Spin2-QMC-1998},
in the zero field limit.

Here we demonstrate its validity for
nonzero magnetic field $H$, which is relevant for ESR,
by discussing
the magnetization and its dependence on $H$ and $T$.
In a classical system with U(1) symmetry around $H$
(thus with $h=0$),
we find the following interesting identity
\begin{equation}
M =
\frac {H}T \int dx \; dy\, \langle L^x(x) L^x(y) \rangle_{\mathrm{cl}},
\label{LL2L}
\end{equation}
where $M=\langle L^z_{\mathrm T} \rangle_{\mathrm{cl}}$ is the
classical uniform magnetization.
This identity is valid for any $H$ and $T$. 
The proof of \eqref{LL2L} is given in the Appendix~\ref{app:LL2L}.
We can easily confirm that this classical magnetization is 
approximately equal to
the uniform magnetization of the corresponding quantum chain.

Since we approximate the renormalization group equation at
the lowest order, the uniform field $\bm H$ and the staggered
field $\bm h$ do not affect the renormalization of the coupling
constant $g$, except that $H$ gives the energy scale $E$ for the
renormalization if $H>T$.
At small but finite field, the uniform susceptibility has
a small deviation from \eqref{chi_uperp2chi_u}, which is
proportional to $H^2$.
We expand the uniform magnetization
$M$ with respect to small $H$:
\begin{align}
\frac{M}L
 &\approx \chi_{u\perp} H \biggl[
  \frac 23
 + \frac{\chi_{u\perp} H^2}2 \int dx \langle
 \bigl( \bm n^\perp(x)\bigr)^2\bigl( \bm n^\perp(0)\bigr)^2
 \rangle_0  \biggr]
 \notag
\end{align}
$\langle \cdot \rangle_0$ denotes the thermal average by the classical
zero-field Hamiltonian \eqref{CHeisenbergH_n}.
The second term can be calculated by using the following formula for
four-point function,
\begin{align}
 &\langle n^a(x) n^b(x) n^c(0) n^d(0) \rangle_{\mathrm{cl}} \notag \\
 &= \frac 19 \delta^{ab}\delta^{cd}
 + \frac{e^{-3|x|/\xi}}{45} \bigl(
 3(\delta^{ac} \delta^{bd} + \delta^{ad} \delta^{bc})
 -2\delta^{ab} \delta^{cd}\bigr),
 \notag
\end{align}
and the result is,
\begin{align}
 \frac ML
 & = \chi_{u\perp} H \biggl[ \frac 23 +
 \frac{4}{135}
 \frac{\xi \chi_{u\perp}H^2}T\biggr] + O(H^5).
 \label{Lz_2ndH}
\end{align}

The uniform (differential) susceptibility is given by
\begin{align}
 \chi_u &=  \frac 23 \chi_{u\perp} \biggl[ 1+\frac 2{15}
 \frac{\chi_{u\perp}\xi H^2}T\biggr]
 \label{chi_u_2ndH} \\
 &= \frac 1{3\pi c} \ln \biggl( \frac{\max{(T,H)}}{\Delta} \biggr)
\biggl[  1+ \frac {cH^2}{15\pi T^2}
\ln^2 \biggl( \frac{\max{(T,H)}}{\Delta}  \biggr) \biggr].
 \label{chi_u_2ndH_v2}
\end{align}
We employ the energy scale $E=\max{(T,H)}$ in the log
correction in \eqref{Lz_2ndH} and \eqref{chi_u_2ndH_v2} instead of
$E=T$ as in the previous Section, because
the magnetic field can exceed the temperature
when we do not consider ESR.
If $H > T$, we should take a cut-off scale as $E=H$.

We have calculated the magnetization density
in the quantum $S=2$ antiferromagnetic chain by quantum
Monte Carlo simulation,
and also in the effective CRM
by classical Monte Carlo (CMC) simulation.
Our method of CMC simulations
of the O(3) CRM is explained in Appendix~\ref{sec.numerics}.
The quantum Monte Carlo simulation was done
using the codes provided by ALPS project\cite{alps05,alps07}.
In the $S=2$ chain, the magnetization
density is defined simply by
$M/L = \langle S^z_{\mathrm T} \rangle/L$.
The numerical results are compared with the
low-field expansion~\eqref{Lz_2ndH} in Fig. \ref{fig.magQvsC}.

The magnetization density in the quantum $S=2$ chain
agrees well with that in the effective classical O(3) NLSM,
and with eq.~\eqref{Lz_2ndH}.
We note that there is no adjustable parameter in
this comparison, owing to the fact that the magnetization
is a conserved quantity.
For higher field $H$, there is a visible discrepancy
between the numerical results and the
analytical prediction~\eqref{Lz_2ndH}.
This is presumably due to the higher order terms
in $H$ which are ignored in eq.~\eqref{Lz_2ndH},
and not because of breakdown of the classical description.
In fact, the CMC result for the effective classical O(3) NLSM
agrees quite well with the quantum $S=2$ chain,
even at higher field $H$.

In Fig. \ref{fig.subtraction}
we show the difference between the magnetization density calculated by
the CMC simulations
and the low field expansion \eqref{Lz_2ndH}.
The difference between the two results is indeed
proportional to $H^5$,
which is the next order in the expansion.
These results on the magnetization supports
the validity of the classical description
for $S \ge 2$ chain, also in a finite magnetic field $H$.

\begin{center}
 \begin{figure}[htbp]
  \includegraphics[scale=0.35]{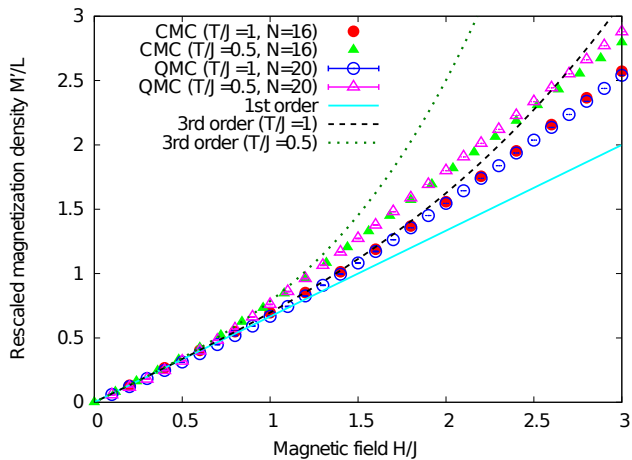}
  \caption{The rescaled magnetization density
  $M/'L= M/L \times 1/(J\chi_{u\perp})$
  for O(3) NLSM is plotted. 
  The open circles and triangles are numerical data
  by classical Monte Carlo simulation.
  The filled circles and triangles with error bars are obtained by quantum
  Monte Carlo simulation.
  We found finite-site effects in QMC data very small, by comparing
  simulations on $20$ sites and $40$ sites.
  We set the unit of the vertical axis so that
  the simple relation $M'/L \approx (2/3) H'$ holds at
  low $H' \equiv H/J <1$.
  The solid line represents the low field expansion
  of $M/L$ up to 1st order of $H$, which is
  $M'/L = M/(LJ\chi_{u\perp})=\frac 23 (H/J)$ by \eqref{Lz_2ndH}.
  The dashed  and dotted lines are 3rd order approximations of
  \eqref{Lz_2ndH} at temperature $T/J=1$ and $T/J=0.5$
  respectively.
  The low $H$ expansion \eqref{Lz_2ndH} is consistent with
  the numerical data in the regime $H/T \lesssim 1$.
  We emphasize that
  CMC and QMC data are consistent
  in any value of the magnetic field.
  }
  \label{fig.magQvsC}
  \end{figure}
 \begin{figure}
  \includegraphics[scale=0.35]{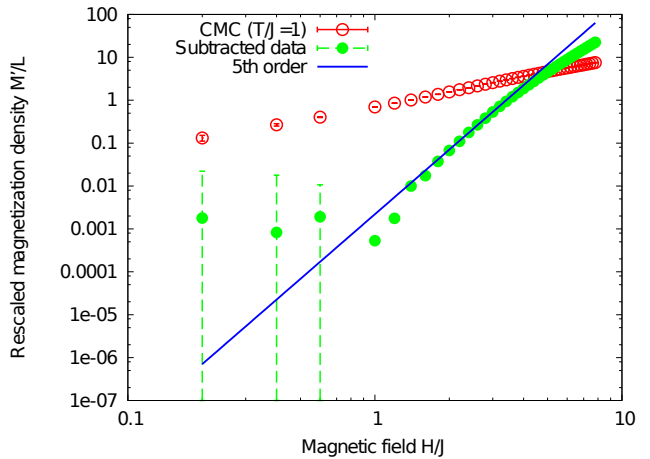}
  \caption{
  The open circles are the classical Monte Carlo data ($T/J=1$)
  which are shown in Fig. \ref{fig.magQvsC}.
  The filled circles are the difference of CMC data and
  \eqref{Lz_2ndH}.
  This subtracted data agree well with the solid line $0.0022\times
  H^5$ whose coefficient is determined by fitting.
  }
  \label{fig.subtraction}
 \end{figure}
\end{center}

\subsection{\label{EOM}Equations of motion}

The classical dynamics of the CRM~\eqref{CNLSMH_lattice}
is studied by solving the classical equation of motion.
In the classical theory,
the commutation relations \eqref{LL=L}, \eqref{Ln=n} and
\eqref{nn=0} are replaced
by the Poisson brackets~\eqref{eq:Poisson}.
These brackets lead to the equations of motion,
\begin{align}
 \frac{d\bm n_j}{dt'} &= (g_R(T) \bm L_j - \bm H') \times \bm n_j,
 \label{dndt'-lattice} \\
 \frac{d\bm L_j}{dt'} &= \frac 1{g_R(T) b^2} \bm n_j \times (\bm n_{j+1} +
 \bm n_{j-1}) \notag \\
 & \qquad \qquad - \bm H' \times \bm L_j - \frac{\Delta'^2_h}{g_R(T)} \bm
 e_x \times \bm n_j .
 \label{dLdt'-lattice}
\end{align}

\section{\label{ESR}ESR spectrum}

In the present approach, the ESR spectrum is obtained from
the classical dynamics of the effective O(3) NLSM theory.
The classical dynamical correlation function is calculated
as follows.
First we generate initial states using the classical Monte
Carlo Method so that the
probability distribution of initial states
is identical to the Boltzmann weight
with the Hamiltonian \eqref{CNLSMH_v2}
at given temperature.
For each initial state,
we solve the equations of motion
\eqref{dndt'-lattice} and \eqref{dLdt'-lattice}
numerically.
We must pay careful attention to 
the total energy, which is a conserved quantity.
Determination of the ESR lineshape
requires the asymptotic, long-time behavior of the
dynamical correlation function.
The time evolution was obtained up to time
$\tprimemax=2000$
with the time step $\delta t' = 0.001$.

Na{\"i}ve numerical integration of the equations of motion
results in violation of the energy conservation.
This makes the scheme unsuitable for ESR calculation,
for which, as we have discussed in the Introduction, 
high accuracy is required.
We applied symplectic methods\cite{Krech98,Landau00}
to assure the conservation law.
We will give detailed explanations
about these numerical methods in Appendix~\ref{sec.numerics}.
The initial state is generated by classical Monte Carlo
simulation of the O(3) NLSM at the given temperature.
\begin{center}
 \begin{figure}[htbp]
  \includegraphics[width=0.4 \textwidth]{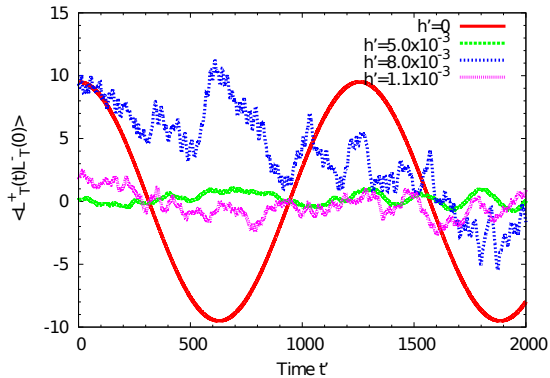}
  \caption{
  Time evolution of $L^+_{\mathrm T}(t) L^-_{\mathrm T}(0)$
  for $S=2$, $T' = 0.15$, and $H' = 0.005$,
  obtained by solving classical equation of motion.
  for one initial state
  generated by the classical Monte Carlo simulation.
  The time evolution appears chaotic for nonzero staggered
  field $h'$, before taking ensemble average.
  }
  \label{fig:time_evolution}
  \end{figure}
\end{center}

An example of the time evolution of
$L^+_{\mathrm T}(t) L^-_{\mathrm T}(0)$
obtained by
numerically solving the equation of motion,
for one initial state is shown in Fig.~\ref{fig:time_evolution}.
For zero staggered field, its exact solution is given by
harmonic oscillation. However, in the presence of a nonvanishing
staggered field, the time evolution looks chaotic.
After Fourier transform, the spectrum is also noisy reflecting
the ``noise'' in the time evolution.

\begin{center}
 \begin{figure}[htbp]
  \includegraphics[width=0.4 \textwidth]{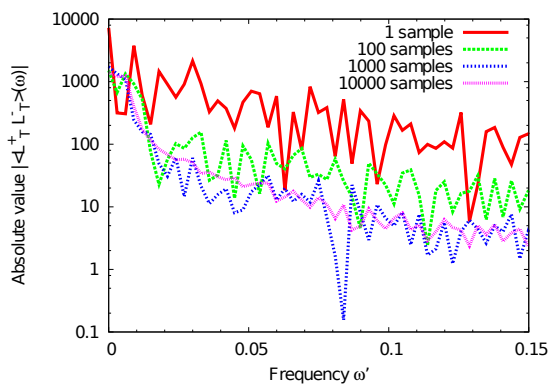}
  \caption{
  The Fourier transform $|\langle L^+_T L^-_T \rangle |(\omega')$
  of the dynamical correlation function~\eqref{eq:LLcorr}
  for $S=2$, $T' = 0.15$, $H' = 0.005$, and
  $h'=0.008$, obtained by averaging over
  various number of sample initial
  states. The approximately white noise is reduced by
  increasing the number of samples.
  }
  \label{fig:ensemble_av}
  \end{figure}
\end{center}

The correlation function~\eqref{eq:LLcorr} is defined by
thermal ensemble average over initial states.
In our calculation,
the above steps are repeated for $10^4$ Monte Carlo samples of initial
states, and the average is taken.
After taking the average, we obtain the dynamical correlation
$\langle L^+_{\mathrm T}(t) L^-_{\mathrm T}(0) \rangle$.
The averaging over many samples of the initial states
reduces the noise in the spectrum.
In Fig.~\ref{fig:ensemble_av} we show the absolute value
of the dynamical correlation function~\eqref{eq:LLcorr}
at $h' = 0.008$,
obtained by average over $1, 100, 1000,$ and $10000$ samples
of initial states.
We note that the data contains both signal and noise.
For $10000$ samples, the result shows small fluctuation
due to noise, around the signal.
For $1$ sample, on the other hand, the observed data
is dominated by the noise, which randomly takes positive
or negative values. (In Fig.~\ref{fig:ensemble_av},
the absolute value is shown.) 

In the range of frequency $0 < \omega' < 0.15$ we are interested,
the power spectrum of the noise is approximately independent of
the frequency $\omega'$; it may be regarded as a white noise.
The reduction of the noise by averaging is clear in the figure,
and the noise is proportional to $1/\sqrt{N_s}$, where $N_s$ is
the number of initial state samples.
However, for a realistic number of samples (we used $N_s \sim 10^4$)
the noise is still not completely negligible.
This effectively constrains resolution in the frequency,
as we discuss below.

Because of the finite time interval $\tprimemax$,
resolution in the frequency space is at most
$2\pi / \tprimemax \sim 0.003$.
It is well known that
a simple Fourier transform of the time dependence
up to the cutoff time $\tprimemax$
leads to an artificial spreading of resonances
known as spectral leakage.~\cite{FT_book}
To suppress the spectral leakage,
a window function is multiplied to the data
before Fourier transform.
Here we apply the Gauss window function.\cite{ChenLandau,EvertzLandau}
Namely, instead of $\langle L^+_{\mathrm T}(t) L^-_{\mathrm T}(0)
\rangle$,
we take the Fourier transformation of
\begin{equation}
 \langle L^+_{\mathrm T}(t) L^-_{\mathrm T}(0) \rangle
  \exp \biggl( -\frac{t^2}{2 \sigma^2}\biggr).
  \label{eq.window}
\end{equation}
Obviously, the width of the Gaussian window $\sigma$ must be
smaller than $\tprimemax$.
Usually the width $\sigma$ is still taken as the same
order as $\tprimemax$, for example
$\sigma \sim 0.4 \tprimemax$.
However, we find that the obtained spectrum is
affected by the noise for such a choice of $\sigma$.
In order to reduce the noise, we take $\sigma$
much smaller than $\tprimemax$.
The reduction of the noise by windowing is discussed
in Appendix~\ref{app:noise_window}.
There is a trade-off between the reduction of noise
 (better for larger $\sigma$) and resolution
in the frequency space (better form smaller $\sigma$).
In this paper, we choose $\sigma = 100 $, which
corresponds to the resolution
$\sigma^{-1} \approx 0.01$ in the frequency space.

\begin{center}
 \begin{figure}[htbp]
  \includegraphics[scale=0.36]{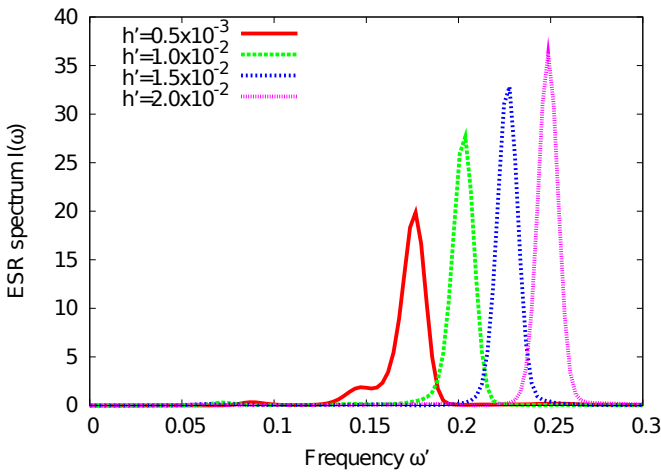}
  \caption{ESR spectra for the $S=10$ case.
  We use dimensionless parameters
  $T' = 0.3, H'= 0.15$.
  While the paramagnetic resonance peak vanishes for $h' \geq 1.5 \times
  10^{-2}$,
  the peak in high frequency side survives.
  The high-frequency peak is much sharper
  compared to the corresponding resonance in the $S=2$ case
  (Fig.~\ref{ESR_T100}).
  }
  \label{fig.ESR_S10}
 \end{figure}
\end{center}

\begin{center}
 \begin{figure}[htbp]
  \includegraphics[scale=0.36]{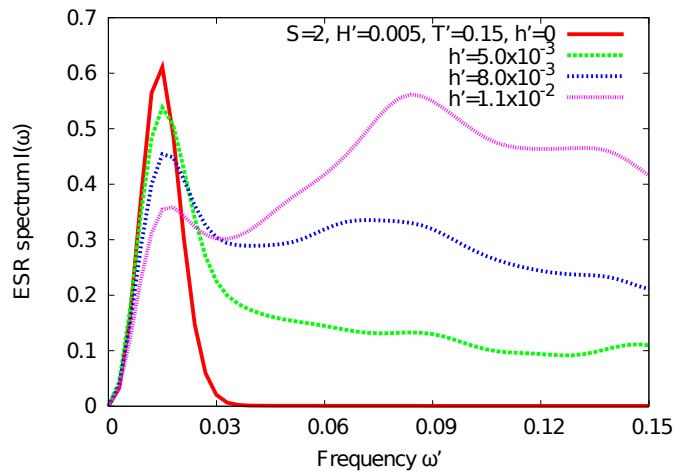}
  \caption{ESR spectra for the dimensionless staggered field
  $h' = h/2JS = 0$, $5.0\times
  10^{-3}$, $8.0 \times 10^{-3}$ and $1.1 \times 10^{-2}$
  are shown.
  Here we assume $S=2$.
  The dimensionless uniform field and temperature
  are respectively given as $H' = H/2JS= 0.005$ and $T' = T/2JS = 0.15$.
  In addition to the paramagnetic resonance ($h'=0$) at $\omega \approx H$,
  we can observe the very broad peak in higher frequency side.}
  \label{ESR_T100}
 \end{figure}
\end{center}

ESR spectra $I(\omega)$ obtained numerically are shown
in Fig. \ref{fig.ESR_S10} for $S=10$, and
in Fig. \ref{ESR_T100} for $S=2$.
They are related to the spectrum
discussed in Fig.~\ref{fig:ensemble_av} via
eqs.~\ref{eq:def_Intensity} and~\ref{chi+-}.
We note that $I(\omega')$
at $\omega' \sim 0$ is suppressed by the factor
${\omega'}^2$ compared
to $|\langle L^+_T L^-_T \rangle|(\omega')$.

The dependence on the spin quantum number $S$
comes only through the effective coupling constant
$g_R$, as given by eqs.~\eqref{renormalizedg} and
\eqref{Delta}.
From Fig.~\ref{fig.ESR_S10} and Fig.~\ref{ESR_T100},
we can see that resonance splits to two peaks as $h$ is increased.
One peak at $\omega \approx H$ results from the paramagnetic resonance.
The intensity of this absorption peak becomes smaller
as the staggered field becomes larger.
For $S=10$, the second peak at higher frequency is sharp,
while the original paramagnetic peak at $\omega' \approx H'$
is almost invisible except for $h' = 0.5 \times 10^{-3}$.
On the other hand, for $S=2$, the original paramagnetic
peak persists and the second peak is very broad for
the studied parameter range.

The second peak at higher frequency $\omega>H$ must be caused
by the staggered magnetic field, because
it becomes dominant while the paramagnetic resonance peak
is suppressed, as we increase the staggered field. 
As we increase $S$,
this peak survives while the paramagnetic resonance
vanishes as shown in Fig. \ref{fig.ESR_S10}.
In the next section,
we will discuss the physical origin of the new peak at $\omega>H$,
as well as the difference between $S=10$ (Fig.~\ref{fig.ESR_S10})
and $S=2$ (Fig.~\ref{ESR_T100}).
The shape of each peak is asymmetric, suggesting
that the lineshape is not a simple Lorentzian, in contrast
to many cases with
strong exchange
interactions.~\cite{KuboTomita54,MoriKawasaki62,Oshikawa99}

We note that the location of the paramagnetic peak in
Fig.~\ref{ESR_T100} is at $\omega' \approx 0.015$ which
is higher than expected $H' = 0.005$.
This is an artifact due to the windowing~\eqref{eq.window},
as discussed in Appendix~\ref{app:noise_window}.


\section{\label{SW}Spin wave theory}

We propose an antiferromagnetic
spin wave theory in order to explain the second peak
at higher frequency observed in Figs.~\ref{fig.ESR_S10}
and~\ref{ESR_T100} .
Here we discuss
linearized fluctuations around the antiferromagnetic
order externally imposed by the staggered field.
In this aspect, it is distinguished from the
standard theory of antiferromagnetic resonance,
in which the antiferromagnetic order is caused by
a spontaneous symmetry breaking.

We initiate our discussion by taking the Lagrangian in
the large $h$ regime, which is \eqref{CNLSML}, 
because the additional peak originate in the large $h$ limit
$h \to + \infty$.
By replacing the bare parameters to the renormalized ones,
we obtain the classical spin wave theory:
\begin{align}
 \mathscr H'_{\mathrm{cl}} &= 
  \dfrac 1{2g_R(T)} \Bigl[(\partial_{t'} \bm m)^2
 + (\partial_x \bm m)^2 \notag \\
 &  \qquad \qquad +  \Delta'^2_y (m^y)^2
  + \Delta'^2_z (m^z)^2\Bigr]
  \label{effectiveCNLSMH}
\end{align}
This Hamiltonian indicates that
the classical spin dynamics in the large $h$ limit is
governed by the two harmonic modes $m^y$ and $m^z$.
These oscillating modes have the eigenfrequencies
\eqref{eq.Deltaz} and \eqref{eq.Deltay}.

Now let us discuss the condition for the spin-wave theory
to be justified.
The spin-wave theory is based on the assumption that the
field $\mathbf{n}$ is polarized along the staggered field
direction, and the fluctuation around the polarized
groundstate is small.
Therefore, the condition can be written as
\begin{equation}
 (m^y)^2 + (m^z)^2 \ll 1.
\label{eq:small_fluctuation}
\end{equation}
Since $\Delta'_z > \Delta'_y$ for any $h'$,
we can expect $\langle (m^z)^2 \rangle < \langle (m^y)^2 \rangle$.
Thus, it is sufficient to require
$\langle (m^y)^2 \rangle \ll 1$.
The Gaussian Hamiltonian \eqref{effectiveCNLSMH} leads
\[
  \langle (m^y)^2 \rangle \approx g_R(T)T' \frac 1{2\Delta'_y}
  \ll 1
\]
and this immediately results in
\begin{equation}
 T' \ll \frac 2{g_R(T)} \Delta'_h
  \label{eq.SW-Trange}
\end{equation}
In terms of the physical staggered field $h$,
this condition can be written as eq.~\eqref{eq:h_for_SWT}.

We note that, the classical approach requires~\eqref{Trange1}, namely
\begin{equation}
 T' \gg \operatorname{max}(\Delta', \Delta'_h, H').
\label{eq:T'-range}
\end{equation}
In particular, $T' \gg \Delta'_h$ is required.
In the classical limit $S \to \infty$,
the renormalized coupling constant $g_R(T)$ approaches zero.
This indicates that the temperature range,
where both \eqref{eq.SW-Trange}
and \eqref{eq:T'-range} hold,
becomes larger as we increase the spin quantum number $S$.

In the classical picture, 
ESR corresponds to
precession of magnetic moments.
If there is no anisotropic interaction,
the {\it total} magnetic moment $\bm L_{\mathrm T}$
precesses around the field $\bm H$ with the
frequency $\omega =H$ without any dissipation.
The modes mentioned above
seem to affect the dynamics of $\bm L_{\mathrm T}$
through the equation of motion \eqref{dndt'-lattice}.
The additional peak reflects
the eigenfrequency of $\bm m$ 
which is different from
the paramagnetic resonance frequency $H$.

In order to see this, the following identity on
the dynamical susceptibility $\chi_{+-}(\omega)$
is useful:
\begin{equation}
 \chi_{+-} (\omega) = \frac{2\langle L^z_{\mathrm T} \rangle}
  {\omega - H} -\frac{\langle [\mathcal A, L^-_{\mathrm T}]\rangle}
  {(\omega - H)^2} + \chi_{\mathcal A^\dagger \mathcal A}(\omega),
  \label{identity}
\end{equation}
where $\mathcal A = \Delta'^2_h n^z_{\mathrm T}/g_R(T)$.
This formula is easily obtained by integrating the
left hand side by parts.
First and second terms in the right hand side
have a singularity only at $\omega = H$.
The last term $\chi_{\mathcal A^\dagger \mathcal A}$ in \eqref{identity}
contributes to the additional
singularity in $\chi''_{+-}$.

The dynamical correlation function of $n^z$ can be
easily calculated within the effective spin-wave
Hamiltonian~\eqref{effectiveCNLSMH}.
Thus the spin-wave approximation predicts an additional
resonance at the frequency identical to eq.~\eqref{eq.Deltaz},
\begin{equation}
 \omega' = {\Delta_z}' = \sqrt{{H'}^2+{\Delta'_h}^2} .
\label{eq:RF_SWT}
\end{equation}
Let us compare this with the numerical results in
Figs.~\ref{fig.ESR_S10} and~\ref{ESR_T100}.
We note that, Figs.~\ref{fig.ESR_S10} and~\ref{ESR_T100}
are shown for $\omega' < T'$, where the classical approximation
would work.

\begin{center}
 \begin{figure}[htbp]
  \includegraphics[scale=0.36]{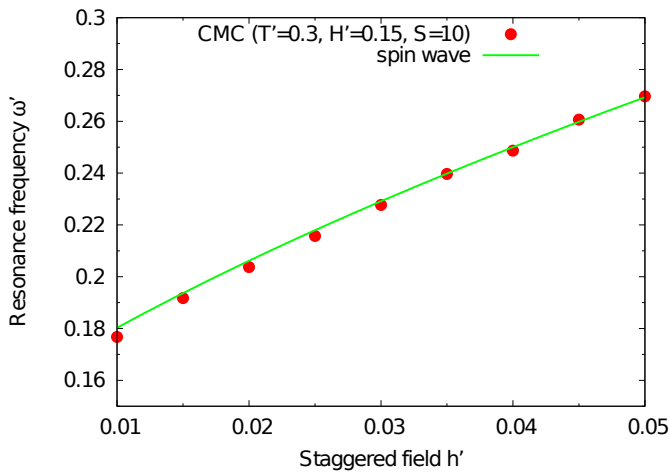}
  \caption{Staggered field dependence of the resonance frequency
  in $S=10$ system at $T'=0.3, H'=0.15$.
  The results agree quite well with the prediction of the spin
  wave theory~\eqref{eq:RF_SWT} (solid line).
  This implies that the resonance peak in the $S=10$ chain
  under a staggered field indeed
  corresponds to the antiferromagnetic spin wave.
  }
  \label{RF_largeh}
 \end{figure}
\end{center}

First we discuss the $S=10$ case, which is highly classical.
The resonance peak observable in Fig.~\ref{fig.ESR_S10}
for non-zero $h'$ can be identified with the antiferromagnetic
spin-wave resonance discussed above, since the original
paramagnetic resonance peak disappears.
The resonance frequencies for several values of $h'$
are plotted in Fig.~\ref{RF_largeh}, and compared
with the theoretical prediction~\eqref{eq:RF_SWT}.
Both conditions~\eqref{eq.SW-Trange} and \eqref{eq:T'-range}
are satisfied in the cases studied in Fig.~\ref{RF_largeh}.
Thus the linear spin-wave approximation of the classical
O(3) NLSM should be valid.
In fact, we find a very good agreement with the
spin-wave theory prediction in this case.

The disappearance of the paramagnetic peak and
the sharp spin-wave resonance corresponding to
eq.~\eqref{eq:RF_SWT} may be understood as
consequences of small fluctuation around the
polarized state along the staggered field.
For a large spin such as $S=10$, the coupling constant
$g_R$ is small and thus eq.~\eqref{eq.SW-Trange}
is easily satisfied, leading to small
fluctuation~\eqref{eq:small_fluctuation}.
Thus the spectrum is well described by the spin-wave
theory which gives the sharp resonance at
frequency~\eqref{eq:RF_SWT}.
The original paramagnetic resonance
at $\omega\sim H$ corresponds to global precession
of spins around the magnetic field $H \parallel z$.
When eq.~\eqref{eq.SW-Trange} holds, the spins are
polarized along the staggered field and thus
the global precession cannot occur;
the paramagnetic resonance is expected to vanish.
This is consistent with the observed behavior
in Fig.~\ref{fig.ESR_S10}.

Next we discuss the $S=2$ case, in which the quantum
fluctuations are stronger corresponding to the
larger value of $g_R$.
In this case, as we have seen in Fig.~\ref{ESR_T100},
two peaks are observed.
The lower frequency peak represents the paramagnetic
resonance at $\omega \sim H$, which is almost independent of
the staggered field.
As in the $S=10$ case,
the higher frequency peak would be identified with
the antiferromagnetic spin-wave.
However, in this case, a quantitative analysis
of the resonance frequency is not possible
because the peak is very broad.
The survival of the paramagnetic peak and the
broadness of the antiferromagnetic spin-wave resonance
for $S=2$ are consequences of large fluctuation owing to
large $g_R$, in contrast to the $S=10$ case discussed above.

We note that, at a lower temperature, the fluctuation
around the polarized state becomes smaller and
the spin-wave theory holds better.
However, for $S=2$, the classical approximation is
no longer valid in this regime.
Nevertheless, the antiferromagnetic spin-wave resonance
should exist also in the quantum regime;
in fact, it exists even in the $S=1$
chain with a staggered field
at low temperature\cite{SakaiShiba94,Mitra94}.
Our classical calculation describes the broadening
of the antiferromagnetic resonance
at higher temperatures.



Finally, we note that, our theory predicts the
spin-wave resonance frequency~\eqref{eq.Deltaz} 
to be independent of temperature.
This is also in agreement with our numerical calculations
(not shown).
While the temperature-independence of the resonance frequency
might seem obvious, it is owing to the lack of the
renormalization of the staggered field for
$N=3$ as shown in eq.~\eqref{hR(T)}.
In a similar analysis based on the O($N$)
NLSM with $N \geq 4$,
the ``spin-wave'' resonance frequency should depend
on the temperature through the logarithmic correction.

\section{\label{summary}Summary and Discussion}

We discussed ESR in the classical limit based on
the O(3) NLSM and the corresponding CRM.
Our discussion is valid in the classical temperature regime,
$\operatorname{max}(\Delta, \Delta_h, H) \ll T \ll JS^2$.
Here the field theoretical picture by O(3) NLSM holds,
and the system behaves classically.
However the microscopic parameters in bare O(3) NLSM are
renormalized by quantum fluctuations.
The dynamics of the O(3) NLSM with renormalized parameters
was then analyzed in the classical limit.
Actual numerical calculations were done for the lattice
version of the NLSM, namely the CRM.

We have demonstrated that numerically calculated ESR
spectra show that
the paramagnetic and 
antiferromagnetic resonance peaks coexist
in this intermediate temperature regime.
The latter is  characteristic of this regime
since it disappears in both lower and higher temperature regions.
We analytically showed that the antiferromagnetic resonance frequency is
$\omega = \sqrt{H^2 + \Delta_h^2}$ based on the
linearized spin wave-theory \eqref{effectiveCNLSMH}.
This agrees well with the numerical results for
a large spin quantum number $S=10$.
For $S=2$, owing to larger fluctuation,
the spin-wave theory is not
quite justified in classical regime.
Nevertheless, a broad resonance is observed in
the spectrum, which is identified with
the (remnant of) antiferromagnetic spin-wave resonance.

It is instructive to compare the present results
with ESR in the $S=1/2$ antiferromagnetic chain.
Although the $S=1/2$ chain at low temperature is described
by a different theoretical approach (bosonization),
as we have discussed in the Introduction,
there is a qualitative similarity between
the two cases.
That is, the paramagnetic resonance at the frequency
$\omega \sim H$ is broadened and eventually disappears
as the temperature is lowered.
On the other hand, at lower temperatures, the
new (antiferromagnetic resonance) peak
at higher frequency becomes dominant.

There is an important difference in the frequency
of the antiferromagnetic resonance.
In both cases, it is given by eq.~\eqref{eq.Deltaz}.
However, the staggered-field-induced gap has different dependence on
the staggered field $h$.
In general, the staggered-field-induced gap is given by
\begin{equation}
\Delta_h \propto h^{\frac{1}{2 - \gamma}},
\end{equation}
where $\gamma$ is the anomalous dimension of the
staggered magnetization.
For the low-energy limit of the $S=1/2$ chain,
the staggered magnetization
has the anomalous dimension $1/2$.
Thus it follows\cite{Oshikawa99} that
\begin{equation}
 \Delta_h \propto h^{2/3},
\end{equation}
up to logarithmic corrections.
In contrast, in the present case, the
anomalous dimension is basically zero and thus
\begin{equation}
 \Delta_h \propto h^{1/2},
\end{equation}
which is consistent with eq.~\eqref{eq.Deltah}.
This difference can be understood as an effect of
strong quantum fluctuations in the $S=1/2$ chain at low temperatures,
on the antiferromagnetic resonance\cite{Nagamiya-AFMR,Keffer-Kittel-AFMR}.

The lineshape of each peak in the present system
appears to be asymmetric and thus non-Lorentzian.
This would also be a significant difference
from the $S=1/2$ case, in which
the broadening of the paramagnetic peak has
Lorentzian form.\cite{Oshikawa99}

In order to apply the present formulation to quantum spin systems,
we need $S \ge 2$ in the case of a single chain.
The temperature range $\operatorname{max}(\Delta, \Delta_h, H)
\ll T \ll JS^2$ is not wide enough for the $S=1$ chains. 
For the $S=2$ Heisenberg antiferromagnetic chain,
we confirmed the validity of the classical O(3) NLSM approach
in a finite magnetic field
by calculating the magnetization density
with quantum Monte Carlo simulations.
The results show a good agreement with the effective
classical O(3) NLSM.

We thus expect that our approach is applicable to the
$S=2$ Haldane chain compound $\mathrm{MnCl_3(bpy)}$,
which may bear an effective staggered field
due to its staggered crystal structure.
It would be interesting to measure ESR spectrum
and compare to our prediction, especially
the appearance of the antiferromagnetic resonance
and its broadening.
The present results can also be applied to spin ladders
or tubes, if the system is described the the classical
O(3) NLSM, and if the staggered field is unfrustrated
as in eq.~\eqref{eq:H_ladder}.

Finally, let us comment on extension of the present
approach to ESR in 2 dimensional antiferromagnets.
The present approach can
be extended to 2 dimensions, as discussed in the Introduction.
The temperature range for the classical approach
is wider in 2 dimensions, as it is not limited from below
by the Haldane gap.
This is favorable for the classical approach
in 2 dimensions.
On the other hand, the increase of the
computational cost in higher dimensions may be
a problem for calculation of ESR spectra which
requires high precision and resolution.
Since ESR in 2 dimensional systems is rather
little understood, it would be interesting
to pursue this approach in 2 dimensions.
We plan to attempt this as a next step.

\section*{Acknowledgement}

This work is supported in part by
Global COE Program ``The Physical Sciences Frontier'', MEXT, Japan
(S.C.F.), JSPS Grant-in-Aid for Scientific Research (KAKENHI)
Nos. 18540341 and 21540381 (S.C.F. and M.O.),
NSERC and CIfAR (I.A.).
Part of the numerical calculations were performed
at the ISSP Supercomputer Center of the University of Tokyo.
We thank Masayuki Hagiwara for useful discussions,
and the ALPS project for providing quantum Monte Carlo
simulation code used in this work.

\appendix

\section{Proof of \eqref{LL2L}}
\label{app:LL2L}

Let us define the partition function
\begin{equation}
 Z(\bm H) = \int \mathscr D\bm n \mathscr D\bm L \prod_x
  \delta(\bm n(x) \cdot  \bm L(x)) \delta(\bm n^2(x) -1)
  e^{-\mathcal H/T}
  \label{PF}
\end{equation}
The key observation is that, thanks to the rotation invariance of the
Hamiltonian, the partition function depends just on the length of
$\bm H$, namely $Z(\bm H) = Z(|\bm H|)$.
Let us consider the two infinitesimal variations of $\bm H$ from
$\bm H_0 = (0, 0, H_0)$.
The first variation is $\bm H = (0, 0, H_0 + \delta H)$.
The partition function can be expanded in terms of $\delta H$ as
\begin{equation}
 Z(\bm H) = Z(H_0) \biggl( 1+ \frac 1T \langle L^z_{\mathrm T} \rangle_0
  \delta H \biggr) + O\bigl( (\delta H)^2 \bigr),
  \label{Z(H)_z}
\end{equation}
up to the first order in $\delta H$.
Here $\langle \cdot \rangle_0$ means that the expectation value with
$\bm H = \bm H_0$.
The second variation we consider is $\bm H = (\delta H', 0, H_0)$.
Again the partition function can be expanded in terms of $\delta H'$.
The first order term actually vanishes here because $\langle
L^x_{\mathrm T} \rangle_0 = 0$ due to the symmetry.
The expansion up to the second order is
\begin{equation}
 Z(\bm H) = Z(H_0) \biggl( 1+ \frac 1{2T^2}
  \langle (L^x_{\mathrm T})^2 \rangle_0 (\delta H')^2 \biggr)
  + O\bigl( (\delta H')^3 \bigr).
  \label{Z(H)_x}
\end{equation}
We compare the variation of $\bm H$ by the two infinitesimal
variation of $\bm H$.
$|\bm H | \approx H_0 + \delta H$ for the first variation and
$|\bm H | \approx H_0 + (\delta H')^2/2H_0$ for the second.
The isotropy of the partition function \eqref{PF} leads that
$\delta H = (\delta H')^2/2H_0$.
Thus, we find the identity
\begin{equation}
 \langle (L^x_{\mathrm T})^2 \rangle_0 = \frac T{H_0}
  \langle L^z_{\mathrm T} \rangle_0,
  \label{LxLx2Lz}
\end{equation}
which is valid for any temperature and for any magnetic field $H_0$.
This result \eqref{LxLx2Lz} is nothing but \eqref{LL2L}.

\section{Numerical methods\label{sec.numerics}}

Our numerical computation of the classical dynamics
consists of two steps:
preparing the initial states at equilibrium, and
solving the equations of motion numerically
from the initial state.

\subsection{Initial states at equilibrium}

The initial states are generated by the
Monte Carlo method as follows.
This method is also used to study static properties 
in Sec.~\ref{magnetization}.
For the sake of simplicity,
we will discuss
in terms of continuum variables for simplicity.
The actual calculation is done for the CRM
on a lattice, for which
$(\partial_x\bm n_j)^2$ is to be replaced by
$- \frac 2{b^2} \bm n_j \cdot \bm n_{j+1}$.

First we eliminate $\bm L$ from
the classical Hamiltonian \eqref{CNLSMH_lattice}.
For simplicity, we denote $\bm n(t=0, x)$ and
$\bm L(t=0, x)$ as $\bm n_0(x)$ and $\bm L_0(x)$
respectively.
The constraint $\bm n_0 \cdot \bm L_0=0$
decrease the degrees of freedom of $\bm L_0$.
For instance, we eliminate $L^z_0$ from the Hamiltonian,
\begin{align*}
 \mathscr H'_{\mathrm{cl}}  &= \frac 1{2g_R(T)} (\partial_x \bm n_0)^2
 + \frac {g_R(T)}{2} \bm L_0^2 -H'L^z_0 -\frac{2h'}{g_R(T)} n^x_0\\
 &= \frac{1}{2g_R(T)} (\partial_x \bm n_0)^2
 + \frac{g_R(T)}{2}\biggl[ (\bm L^\perp_0)^2
 + \frac{(\bm n^\perp_0 \cdot \bm L^\perp_0)^2}{(n^z_0)^2}\biggr]
 \\
 & \qquad + H' \frac{\bm n^\perp_0 \cdot \bm L^\perp_0}{n^z_0}
 - \frac{2h'}{g_R(T)} n^x_0,
\end{align*}
where $\bm L^\perp_0$ and $\bm n^\perp_0$ are defined as
$\bm L^\perp_0 \equiv (L^x_0, L^y_0, 0)$
and $\bm n^\perp_0 = (n^x_0, n^y_0, 0)$
respectively.
$\bm L^\perp_0$ is distributed with Gaussian distribution
with an average
$-\chi_{u\perp} H  \bm n^{\perp}_0 n^z_0$,
and the variance $T\chi_{u\perp}$.
We can integrate out $\bm L^\perp_0$ because
the Hamiltonian is harmonic in $\bm L^\perp_0$.
\begin{equation}
 \mathscr H_{\mathrm{eff}}=
 \frac1{2g_R(T)} (\partial_x \bm n_0)^2
 + \frac{H'^2}{2g_R(T)} (n^z_0)^2 -\frac{2h'}{g_R(T)}n^x_0
\label{eq.effH}
\end{equation}
Starting from an arbitrary configuration of
$\{\bm n(x)\}$,
we thermalize it by classical Monte Carlo method
based on the Wolff algorithm\cite{Wolff89}.
$\bm L^\perp_0 $ is calculated by
$\bm L^\perp_0 = -H' \bm n_0^\perp n^z_0/g_R(T)
+ \bm R$.
 $\bm R$ is a random vector
whose distribution is Gaussian with an  average 0 and
a variance $T\chi_{u\perp}$.
The constraint $\bm n_0 \cdot \bm L_0 = 0$
determines $\langle L^z_0 \rangle$.

\subsection{Calculation of dynamical
correlation functions}

We need to solve the classical equations of motion
\eqref{dndt'-lattice} and \eqref{dLdt'-lattice} numerically
with the initial states produced in Appendix~\ref{sec.numerics}.
We should pay attention to the conserved quantities,
the total energy and the uniform magnetization
(if there is no staggered field) in our system.
Ordinary numerical methods for solving the equations motion,
for instance, Runge-Kutta method or the predictor-corrector method,
result in the violation
of the conservation law.
This violation is caused by discretizations.
Even if the continuum equations of motion preserve
the conservation laws, the discretized version of
the equations of motion
do not necessarily preserve them.

This problem is serious in our situation.
ESR spectrum requires asymptotic behavior of
the dynamical correlation function, namely
$\langle L^+_{\mathrm T}(t) L^-_{\mathrm T}(0) \rangle$
at $t \to +\infty$.
The configuration $\{ \bm n(t,x), \, \bm L(t,x)\}$
may be far from the real one
which exactly conserves the energy
once the conservation law is violated.

\subsubsection{Symplectic methods}

We overcame this difficulty by applying 
``symplectic'' methods \cite{Krech98,Landau00}.
This algorithm is based on the Suzuki-Trotter
decomposition \cite{Suzuki90,Suzuki92} of
integrators.
Applications to classical spin systems
are explained in detail in
Refs.~\onlinecite{Krech98,Landau00}.

Let us consider the time evolution of a scalar
$f(t)$ for simplicity.
We assume the equation of motion for $f(t)$ as
\[
 \frac{df}{dt} = \mathcal F(f(t)) = \mathcal F_1(f(t))
 + \mathcal F_2(f(t)).
\]
Further we assume that
we can exactly solve the equations
\begin{equation}
 \frac{df}{dt} = \mathcal F_i(f)
  \label{eq.EoMfi}
\end{equation}
for $i=1, 2$.
Infinitesimal time evolution of $f$ from
$t$ to $t+\delta t$ has an exponential form
\begin{equation}
 f(t+\delta t) = e^{\delta t \mathfrak h} f(t)
  =e^{\delta t(\mathfrak h_1+ \mathfrak h_2)} f(t).
  \label{eq.EoMf}
\end{equation}
Since we can solve \eqref{eq.EoMfi} exactly,
we know the explicit form of
$\mathfrak h_1$ and $\mathfrak h_2$ and thus,
$e^{\delta t \mathfrak h_1}$ and $e^{\delta t\mathfrak h_2}$.
The essence of the symplectic method is in
approximating the operator $e^{\delta t(\mathfrak h_1+
\mathfrak h_2)}$ by exactly known operators
$e^{\delta t \mathfrak h_1}$ and $e^{\delta t \mathfrak h_2}$.

The fourth order approximation \cite{Yoshida90}
is known:
\begin{equation}
 e^{\delta t (\mathfrak h_1 + \mathfrak h_2)}
  = \prod_{i=1}^5 e^{p_i \delta t \mathfrak h_1/2}
  e^{p_i\delta t \mathfrak h_2} e^{p_i \delta t
  \mathfrak h_1/2},
  \label{eq.4thSD}
\end{equation}
where $p_1=p_2=p_4=p_5 = 1/(4-4^{1/3})$ and
$p_3=-4^{1/3}/(4-4^{1/3})$.
We can extend the decomposition \eqref{eq.4thSD}
to $2n$th order one with $n \ge 3$.
By applying Baker-Campbell-Hausdorff (BCH) formula
repeatedly, we can show that the right hand side
in \eqref{eq.4thSD} is
$e^{\delta t \mathfrak h + \mathcal O((\delta t)^6)}$.
It is easily proved that $2n$-th order decomposition
of this type is $(2n+1)$-th order approximation.
It follows immediately that the symplectic method
is more accurate than other naive methods and that
the conserved quantities
during the calculation do not
deviate from the exact value.
Let us discuss a specific example about the latter.
In Hamilton systems, the operator $\mathfrak h$
is the Hamiltonian itself.
Let us write
the approximated $\mathfrak h$
in \eqref{eq.4thSD}
as $\mathfrak h_4$.
The difference $\mathfrak h_4-\mathfrak h$ 
is bounded as
\begin{equation}
 \mathfrak h_4- \mathfrak h = \mathrm{const.}
  \times (\delta t)^5
  \label{eq.accuracy}
\end{equation} 
according to \eqref{eq.4thSD} and BCH formula.
\eqref{eq.accuracy} is much more accurate than that
of 4th order Runge-Kutta method.
Let us consider the total energy at time $t$
and denote it as $E(t)$.
First we prepare initial state and fix $E(t=0)$.
By applying the symplectic method, $E(t)-E(0)$
is order of $(\delta t)^5$, thus, $E(t)-E(0) \sim (\delta t)^5$.
However, by the Runge-Kutta method,
the error is
increasing at least linearly with respect to $t$.
Thus, $E(t)-E(0) \sim t$ becomes quite large
if we want to consider large $t$.

Thus, if we apply the symplectic method,
the total energy conserves with high accuracy
during the discretized time evolution.

\subsubsection{Application to classical rotor model}

The application of this symplectic method
is as follows.
We decompose the equations of motion \eqref{dndt'-lattice}
and \eqref{dLdt'-lattice}:
\begin{subequations}
 \begin{align}
  &\left\{
   \begin{array}{l}
    \partial_{t'} \bm n_j = g_R(T) \bm L_j \times
     \bm n_j \\[+8pt]
    \partial_{t'} \bm L_j=0
   \end{array}
  \right.
  \label{eq.dEoM1}
   \\[+16pt]
  &\left\{
   \begin{array}{l}
    \partial_{t'} \bm n_j = -\bm H' \times
     \bm n_j \\
     [+10pt]
    \partial_{t'} \bm L_j= - \bm H' \times \bm n_j
   \end{array}
  \right.
  \label{eq.dEoM2}
 \\[+16pt]
  &\left\{
   \begin{array}{l}
    \partial_{t'} \bm n_j = 0 \\
     [+8pt]
    \partial_{t'} \bm L_j= 
    \dfrac 1{b^2g_R(T)}\bm n_j \times
    (\bm n_{j+1} + \bm n_{j-1})
    \dfrac{2h'}{g_R(T)} \bm e_x \times \bm n_j
   \end{array}
  \right.
  \label{eq.dEoM3}
\end{align}
\end{subequations}
Each of these equations are exactly solvable.
For instance, \eqref{eq.dEoM1} is solved as follows:
\begin{align*}
 \bm n_j(t'+\delta t')&= \bm n_j(t')
 \cos \Bigl(g_R(T)\delta t' \,|\bm L_j(t)|\Bigr) \\
 & \quad + \frac{\bm L_j(t') \times \bm n_j(t')}
 {|\bm L_j(t')|} \sin
 \Bigl(g_R(T)\delta t' \,|\bm L_j(t)|\Bigr)\\
\bm L_j(t'+\delta t') &= \bm L_j(t')
\end{align*}
We consider this step from $t'$ to $t'+\delta t'$
as $e^{\delta t' A}$.
In the same way we define the time step $e^{\delta t'B}$
and $e^{\delta t' C}$ as
\begin{align*}
 \bm n_j(t'+\delta t') &= \bm n_j(t') \cos (H\delta t') - \frac{\bm H' \times
 \bm n_j(t')}{H'} \sin (H\delta t') \\
 \bm L_j(t'+\delta t')&= \bm L_j(t') \cos(H\delta t') - \frac{\bm H' \times
 \bm L_j(t')}{H'} \sin (H\delta t')
\end{align*}
and
\begin{align*}
 \bm n_j(t'+\delta t')&= \bm n_j(t') \\
 \bm L_j(t'+\delta t')&= \bm L_j(t') 
 + \delta t' \biggl(
 \frac 1{g_R(T)} \bm n_j(t')  \\
 & \> \times (\bm n_{j+1}(t') + \bm n_{j-1}
 (t'))
 -\frac{2h'}{g_R(T)} \bm e_x \times \bm n_j(t')\biggr)
\end{align*}
respectively.
Actual time evolution is considered as $e^{\delta t' (A+B+C)}$.
Second order decomposition of $e^{\delta t'(A+B+C)}$ is
$ e^{\delta t'(A+B+C)} = e^{\delta t' A/2}e^{\delta t' B/2}e^{\delta t' C}
 e^{\delta t' B/2} e^{\delta t' C/2}$.
In this manner, we can apply these to the symplectic method.
Fourth order calculation conserve the total energy
within the precision $\Delta E/E \lesssim 10^{-8}$.

\section{Effects of windowing}
\label{app:noise_window}

In this Appendix, we discuss reduction of white noise by
windowing. For simplicity, here we identify
time $t$ and frequency $\omega$ with the rescaled
variables $t'$ and $\omega'$.

In general, the Fourier transform of a real-time function $f(t)$ is
defined by
\begin{equation}
 F(\omega) = \int dt \; f(t) e^{i \omega t}
\end{equation}
From the definition, Parseval's identity
\begin{equation}
 \int \frac{d\omega}{2\pi} \; |F(\omega)|^2 = \int dt\; |f(t)|^2,
\label{eq:Parseval}
\end{equation}
follows.
In practice, we use $f(t)$ at discrete time with time step
$\Delta t$ and upper bound $\tmax$.
In this case, the integrals are replaced by discrete sums
\begin{align}
 \int dt & \rightarrow \Delta t
\sum_{0 \leq n < \tmax/\Delta t},  \\
 \int \frac{d\omega}{2\pi}
& \rightarrow \frac{1}{\tmax}
\sum_{0 \leq m < \tmax/\Delta t} .
\end{align}
Assuming white noise, we consider the case
$|\fnoise(t)|^2 = \delta^2$ at any discrete time and
$|\Fnoise(\omega)|^2$ is independent of frequency $\omega$.
Then Parseval's identity implies
\begin{equation}
 |\Fnoise(\omega)|^2 = \delta^2 \tmax \Delta t  .
\end{equation}
Namely, the white noise in the frequency space
increases as the sampling time interval
$\tmax$ is increased.
The white noise could be reduced simply
by taking smaller interval $\tmax$,
but it enhances spectral leakage.
A better alternative is to use the window function
with smaller width, effectively reducing the sampling
time interval.
In this paper, we use the Gaussian window function.
The Fourier transform with the windowing is defined as
\begin{equation}
 \tilde{F}(\omega) \equiv
  \int dt \; f(t) e^{-t^2/(2 \sigma^2)} e^{i\omega t} .
\end{equation}
The couterpart of eq.~\eqref{eq:Parseval} becomes
\begin{equation}
 \int \frac{d\omega}{2\pi} |\tilde{F} (\omega)|^2
= \int dt \; |f(t)|^2 e^{-t^2/\sigma^2} .
\end{equation}
For the white noise with $|\fnoise(t)|^2 = \delta^2$, and
we find
\begin{equation}
 |\tildeFnoise(\omega)|^2 = \delta^2 \sqrt{\pi} \sigma \Delta t .
\end{equation}
In our study, $\fnoise(t)$ corresponds to the noise in the
dynamical correlation function
$\langle L^+_{\mathrm T}(t) L^-_{\mathrm T}(0) \rangle$.
sampling time step is
$\Delta t \sim 0.1$, and $\delta \sim 0.02$.
For $\sigma = 100$, this gives
$|\tildeFnoise(\omega)| \sim 0.2$,
which is about 2 \% of the observed signal.
In other words, this signal-to-noise ratio is achieved
with $\sigma = 100$.

On the other hand, there is a side-effect of the
windowing.
In the absence of the staggered field $h'=0$, the
exact spectrum is proportional to $\delta(\omega-H)$,
according to the general result in the absence of anisotropy.
However, in Fig.~\ref{ESR_T100}, the paramagnetic resonance
appears to have a finite width and the peak seems
to be shifted to higher frequency $\omega' \approx 0.015$
from the expected $\omega' = H' = 0.005$.

This is actually due to the the artificial
linewidth $\sigma^{-1}$ introduced through the window
function \eqref{eq.window}.
It actually dominates the
linewidth of the paramagnetic peak in Fig.~\ref{ESR_T100}.
Once a finite width is induced, the peak in the
dynamical susceptibility is distorted by the factor
$1-e^{-\omega/T}$ (which is proportional to $\omega$
at low frequency and give more weights to higher
frequencies) in the definition, eq.~\eqref{chi+-}.
For a uniform field $H'$ sufficiently larger than the
resolution, we find the (resolution-limited) paramagnetic
resonance peak at $\omega' \sim H'$, as expected.
In Fig.~\ref{ESR_T100}, we have shown the data for
small $H'$ so that the additional broad peak can
be observable in a wide range of frequency within
the validity of the classical approximation.
The low-frequency peak in
Fig.~\ref{ESR_T100} is smoothly connected to
the paramagnetic resonance peak at $\omega' \sim H'$,
when the uniform magnetic field $H'$ is increased;
thus it can be identified with the paramagnetic
resonance peak despite the apparent width and shift
due to the frequency resolution in the calculation.
In fact, the paramagnetic peak itself can be observed
with a larger value of $\sigma$.


\begin{thebibliography}{37}
\expandafter\ifx\csname natexlab\endcsname\relax\def\natexlab#1{#1}\fi
\expandafter\ifx\csname bibnamefont\endcsname\relax
  \def\bibnamefont#1{#1}\fi
\expandafter\ifx\csname bibfnamefont\endcsname\relax
  \def\bibfnamefont#1{#1}\fi
\expandafter\ifx\csname citenamefont\endcsname\relax
  \def\citenamefont#1{#1}\fi
\expandafter\ifx\csname url\endcsname\relax
  \def\url#1{\texttt{#1}}\fi
\expandafter\ifx\csname urlprefix\endcsname\relax\def\urlprefix{URL }\fi
\providecommand{\bibinfo}[2]{#2}
\providecommand{\eprint}[2][]{\url{#2}}

\bibitem[{\citenamefont{Kubo and Tomita}(1954)}]{KuboTomita54}
\bibinfo{author}{\bibfnamefont{R.}~\bibnamefont{Kubo}} \bibnamefont{and}
  \bibinfo{author}{\bibfnamefont{K.}~\bibnamefont{Tomita}},
  \bibinfo{journal}{J. Phys. Soc. Jpn} \textbf{\bibinfo{volume}{9}},
  \bibinfo{pages}{888} (\bibinfo{year}{1954}).

\bibitem[{\citenamefont{Mori and Kawasaki}(1962)}]{MoriKawasaki62}
\bibinfo{author}{\bibfnamefont{H.}~\bibnamefont{Mori}} \bibnamefont{and}
  \bibinfo{author}{\bibfnamefont{K.}~\bibnamefont{Kawasaki}},
  \bibinfo{journal}{Prog. Theor. Phys.} \textbf{\bibinfo{volume}{28}},
  \bibinfo{pages}{971} (\bibinfo{year}{1962}).

\bibitem[{\citenamefont{Chakravarty et~al.}(1989)\citenamefont{Chakravarty,
  Halperin, and Nelson}}]{CHN}
\bibinfo{author}{\bibfnamefont{S.}~\bibnamefont{Chakravarty}},
  \bibinfo{author}{\bibfnamefont{B.~I.} \bibnamefont{Halperin}},
  \bibnamefont{and} \bibinfo{author}{\bibfnamefont{D.~R.}
  \bibnamefont{Nelson}}, \bibinfo{journal}{Phys. Rev. B}
  \textbf{\bibinfo{volume}{39}}, \bibinfo{pages}{2344} (\bibinfo{year}{1989}).

\bibitem[{\citenamefont{Ty\ifmmode~\check{c}\else \v{c}\fi{}
  et~al.}(1989)\citenamefont{Ty\ifmmode~\check{c}\else \v{c}\fi{}, Halperin,
  and Chakravarty}}]{TycHalperinChakravarty-PRL1989}
\bibinfo{author}{\bibfnamefont{S.}~\bibnamefont{Ty\ifmmode~\check{c}\else
  \v{c}\fi{}}}, \bibinfo{author}{\bibfnamefont{B.~I.} \bibnamefont{Halperin}},
  \bibnamefont{and}
  \bibinfo{author}{\bibfnamefont{S.}~\bibnamefont{Chakravarty}},
  \bibinfo{journal}{Phys. Rev. Lett.} \textbf{\bibinfo{volume}{62}},
  \bibinfo{pages}{835} (\bibinfo{year}{1989}).

\bibitem[{\citenamefont{Damle and Sachdev}(1998)}]{DamleSachdev98}
\bibinfo{author}{\bibfnamefont{K.}~\bibnamefont{Damle}} \bibnamefont{and}
  \bibinfo{author}{\bibfnamefont{S.}~\bibnamefont{Sachdev}},
  \bibinfo{journal}{Phys. Rev. B} \textbf{\bibinfo{volume}{57}},
  \bibinfo{pages}{8307} (\bibinfo{year}{1998}).

\bibitem[{\citenamefont{Buragohain and Sachdev}(1999)}]{BuragohainSachdev99}
\bibinfo{author}{\bibfnamefont{C.}~\bibnamefont{Buragohain}} \bibnamefont{and}
  \bibinfo{author}{\bibfnamefont{S.}~\bibnamefont{Sachdev}},
  \bibinfo{journal}{Phys. Rev. B} \textbf{\bibinfo{volume}{59}},
  \bibinfo{pages}{9285} (\bibinfo{year}{1999}).

\bibitem[{\citenamefont{Affleck}(1990)}]{Affleck90}
\bibinfo{author}{\bibfnamefont{I.}~\bibnamefont{Affleck}},
  \bibinfo{journal}{Phys. Rev. B} \textbf{\bibinfo{volume}{41}},
  \bibinfo{pages}{6697} (\bibinfo{year}{1990}).

\bibitem[{\citenamefont{Huang and Affleck}(2004)}]{HuangAffleck04}
\bibinfo{author}{\bibfnamefont{H.}~\bibnamefont{Huang}} \bibnamefont{and}
  \bibinfo{author}{\bibfnamefont{I.}~\bibnamefont{Affleck}},
  \bibinfo{journal}{Phys. Rev. B} \textbf{\bibinfo{volume}{69}},
  \bibinfo{pages}{184414} (\bibinfo{year}{2004}).

\bibitem[{\citenamefont{Todo and Kato}(2001)}]{Todo01}
\bibinfo{author}{\bibfnamefont{S.}~\bibnamefont{Todo}} \bibnamefont{and}
  \bibinfo{author}{\bibfnamefont{K.}~\bibnamefont{Kato}},
  \bibinfo{journal}{Phys. Rev. Lett.} \textbf{\bibinfo{volume}{87}},
  \bibinfo{pages}{047203} (\bibinfo{year}{2001}).

\bibitem[{\citenamefont{Sakai and Shiba}(1994)}]{SakaiShiba94}
\bibinfo{author}{\bibfnamefont{T.}~\bibnamefont{Sakai}} \bibnamefont{and}
  \bibinfo{author}{\bibfnamefont{H.}~\bibnamefont{Shiba}}, \bibinfo{journal}{J.
  Phys. Soc. Jpn} \textbf{\bibinfo{volume}{63}}, \bibinfo{pages}{867}
  (\bibinfo{year}{1994}).

\bibitem[{\citenamefont{Mitra and Halperin}(1994)}]{Mitra94}
\bibinfo{author}{\bibfnamefont{P.~P.} \bibnamefont{Mitra}} \bibnamefont{and}
  \bibinfo{author}{\bibfnamefont{B.~I.} \bibnamefont{Halperin}},
  \bibinfo{journal}{Phys. Rev. Lett.} \textbf{\bibinfo{volume}{72}},
  \bibinfo{pages}{912} (\bibinfo{year}{1994}).

\bibitem[{\citenamefont{Oshikawa and Affleck}(1997)}]{Oshikawa97}
\bibinfo{author}{\bibfnamefont{M.}~\bibnamefont{Oshikawa}} \bibnamefont{and}
  \bibinfo{author}{\bibfnamefont{I.}~\bibnamefont{Affleck}},
  \bibinfo{journal}{Phys. Rev. Lett.} \textbf{\bibinfo{volume}{79}},
  \bibinfo{pages}{2883} (\bibinfo{year}{1997}).

\bibitem[{\citenamefont{Meyer et~al.}(1982)\citenamefont{Meyer, Gleizes,
  Girerd, Verdaguer, and Kahn}}]{Meyer82}
\bibinfo{author}{\bibfnamefont{A.}~\bibnamefont{Meyer}},
  \bibinfo{author}{\bibfnamefont{A.}~\bibnamefont{Gleizes}},
  \bibinfo{author}{\bibfnamefont{J.}~\bibnamefont{Girerd}},
  \bibinfo{author}{\bibfnamefont{M.}~\bibnamefont{Verdaguer}},
  \bibnamefont{and} \bibinfo{author}{\bibfnamefont{O.}~\bibnamefont{Kahn}},
  \bibinfo{journal}{Inorg. Chem} \textbf{\bibinfo{volume}{21}},
  \bibinfo{pages}{1729} (\bibinfo{year}{1982}).

\bibitem[{\citenamefont{Date et~al.}(1970)\citenamefont{Date, Yamazaki,
  Motokawa, and Tazawa}}]{Date-Cubenz}
\bibinfo{author}{\bibfnamefont{M.}~\bibnamefont{Date}},
  \bibinfo{author}{\bibfnamefont{H.}~\bibnamefont{Yamazaki}},
  \bibinfo{author}{\bibfnamefont{M.}~\bibnamefont{Motokawa}}, \bibnamefont{and}
  \bibinfo{author}{\bibfnamefont{S.}~\bibnamefont{Tazawa}},
  \bibinfo{journal}{Progress of Theoretical Physics Supplement}
  \textbf{\bibinfo{volume}{46}}, \bibinfo{pages}{194} (\bibinfo{year}{1970}).

\bibitem[{\citenamefont{Oshikawa and Affleck}(1999)}]{Oshikawa99}
\bibinfo{author}{\bibfnamefont{M.}~\bibnamefont{Oshikawa}} \bibnamefont{and}
  \bibinfo{author}{\bibfnamefont{I.}~\bibnamefont{Affleck}},
  \bibinfo{journal}{Phys. Rev. Lett.} \textbf{\bibinfo{volume}{82}},
  \bibinfo{pages}{5136} (\bibinfo{year}{1999}).

\bibitem[{\citenamefont{Oshikawa and Affleck}(2002)}]{OshikawaAffleck02}
\bibinfo{author}{\bibfnamefont{M.}~\bibnamefont{Oshikawa}} \bibnamefont{and}
  \bibinfo{author}{\bibfnamefont{I.}~\bibnamefont{Affleck}},
  \bibinfo{journal}{Phys. Rev. B} \textbf{\bibinfo{volume}{65}},
  \bibinfo{pages}{134410} (\bibinfo{year}{2002}).

\bibitem[{\citenamefont{Granroth et~al.}(1996)\citenamefont{Granroth, Meisel,
  Chaparala, Jolic{\oe}ur, Ward, and Talham}}]{Granroth96}
\bibinfo{author}{\bibfnamefont{G.~E.} \bibnamefont{Granroth}},
  \bibinfo{author}{\bibfnamefont{M.~W.} \bibnamefont{Meisel}},
  \bibinfo{author}{\bibfnamefont{M.}~\bibnamefont{Chaparala}},
  \bibinfo{author}{\bibfnamefont{T.}~\bibnamefont{Jolic\oe{}ur}},
  \bibinfo{author}{\bibfnamefont{B.~H.} \bibnamefont{Ward}}, \bibnamefont{and}
  \bibinfo{author}{\bibfnamefont{D.~R.} \bibnamefont{Talham}},
  \bibinfo{journal}{Phys. Rev. Lett.} \textbf{\bibinfo{volume}{77}},
  \bibinfo{pages}{1616} (\bibinfo{year}{1996}).

\bibitem[{\citenamefont{Sato and Oshikawa}(2004)}]{Sato-coupled-staggered}
\bibinfo{author}{\bibfnamefont{M.}~\bibnamefont{Sato}} \bibnamefont{and}
  \bibinfo{author}{\bibfnamefont{M.}~\bibnamefont{Oshikawa}},
  \bibinfo{journal}{Phys. Rev. B} \textbf{\bibinfo{volume}{69}},
  \bibinfo{pages}{054406} (\bibinfo{year}{2004}).

\bibitem[{\citenamefont{Zhao et~al.}(2006)\citenamefont{Zhao, Wang, Xiang, Su,
  Yu, Lou, and Chen}}]{Zhao-ladder-staggered}
\bibinfo{author}{\bibfnamefont{J.}~\bibnamefont{Zhao}},
  \bibinfo{author}{\bibfnamefont{X.}~\bibnamefont{Wang}},
  \bibinfo{author}{\bibfnamefont{T.}~\bibnamefont{Xiang}},
  \bibinfo{author}{\bibfnamefont{Z.}~\bibnamefont{Su}},
  \bibinfo{author}{\bibfnamefont{L.}~\bibnamefont{Yu}},
  \bibinfo{author}{\bibfnamefont{J.}~\bibnamefont{Lou}}, \bibnamefont{and}
  \bibinfo{author}{\bibfnamefont{C.}~\bibnamefont{Chen}},
  \bibinfo{journal}{Phys. Rev. B} \textbf{\bibinfo{volume}{73}},
  \bibinfo{pages}{012411} (\bibinfo{year}{2006}).

\bibitem[{\citenamefont{Nagamiya}(1951)}]{Nagamiya-AFMR}
\bibinfo{author}{\bibfnamefont{T.}~\bibnamefont{Nagamiya}},
  \bibinfo{journal}{Progress of Theoretical Physics}
  \textbf{\bibinfo{volume}{6}}, \bibinfo{pages}{342} (\bibinfo{year}{1951}).

\bibitem[{\citenamefont{Keffer and Kittel}(1952)}]{Keffer-Kittel-AFMR}
\bibinfo{author}{\bibfnamefont{F.}~\bibnamefont{Keffer}} \bibnamefont{and}
  \bibinfo{author}{\bibfnamefont{C.}~\bibnamefont{Kittel}},
  \bibinfo{journal}{Phys. Rev.} \textbf{\bibinfo{volume}{85}},
  \bibinfo{pages}{329} (\bibinfo{year}{1952}).

\bibitem[{\citenamefont{Iitaka and Ebisuzaki}(2003)}]{Iitaka-ESR}
\bibinfo{author}{\bibfnamefont{T.}~\bibnamefont{Iitaka}} \bibnamefont{and}
  \bibinfo{author}{\bibfnamefont{T.}~\bibnamefont{Ebisuzaki}},
  \bibinfo{journal}{Phys. Rev. Lett.} \textbf{\bibinfo{volume}{90}},
  \bibinfo{pages}{047203} (\bibinfo{year}{2003}).

\bibitem[{\citenamefont{Br{\'e}zin and Zinn-Justin}(1976)}]{BZJ76}
\bibinfo{author}{\bibfnamefont{E.}~\bibnamefont{Br{\'e}zin}} \bibnamefont{and}
  \bibinfo{author}{\bibfnamefont{J.}~\bibnamefont{Zinn-Justin}},
  \bibinfo{journal}{Phys. Rev. B} \textbf{\bibinfo{volume}{14}},
  \bibinfo{pages}{3110} (\bibinfo{year}{1976}).

\bibitem[{\citenamefont{Affleck et~al.}(1989)\citenamefont{Affleck, Gepner,
  Schulz, and Ziman}}]{Affleck89}
\bibinfo{author}{\bibfnamefont{I.}~\bibnamefont{Affleck}},
  \bibinfo{author}{\bibfnamefont{D.}~\bibnamefont{Gepner}},
  \bibinfo{author}{\bibfnamefont{H.~J.} \bibnamefont{Schulz}},
  \bibnamefont{and} \bibinfo{author}{\bibfnamefont{T.}~\bibnamefont{Ziman}},
  \bibinfo{journal}{J. Phys. A:Math. Gen.} \textbf{\bibinfo{volume}{22}},
  \bibinfo{pages}{511} (\bibinfo{year}{1989}).

\bibitem[{\citenamefont{Fisher}(1964)}]{Fisher-classical}
\bibinfo{author}{\bibfnamefont{M.~E.} \bibnamefont{Fisher}},
  \bibinfo{journal}{Am. J. Phys.} \textbf{\bibinfo{volume}{32}},
  \bibinfo{pages}{343} (\bibinfo{year}{1964}).

\bibitem[{\citenamefont{Kim et~al.}(1998)\citenamefont{Kim, Greven, Wiese, and
  Birgeneau}}]{YJKim-Spin2-QMC-1998}
\bibinfo{author}{\bibfnamefont{Y.}~\bibnamefont{Kim}},
  \bibinfo{author}{\bibfnamefont{M.}~\bibnamefont{Greven}},
  \bibinfo{author}{\bibfnamefont{U.-J.} \bibnamefont{Wiese}}, \bibnamefont{and}
  \bibinfo{author}{\bibfnamefont{R.}~\bibnamefont{Birgeneau}},
  \bibinfo{journal}{Eur. Phys. J. B} \textbf{\bibinfo{volume}{4}},
  \bibinfo{pages}{291} (\bibinfo{year}{1998}).

\bibitem[{\citenamefont{Alet et~al.}(2005)\citenamefont{Alet, Dayal, Grzesik,
  Honecker, Koerner, Laeuchli, Manmana, MsCulloch, Michel, Noack
  et~al.}}]{alps05}
\bibinfo{author}{\bibfnamefont{F.}~\bibnamefont{Alet}},
  \bibinfo{author}{\bibfnamefont{P.}~\bibnamefont{Dayal}},
  \bibinfo{author}{\bibfnamefont{A.}~\bibnamefont{Grzesik}},
  \bibinfo{author}{\bibfnamefont{A.}~\bibnamefont{Honecker}},
  \bibinfo{author}{\bibfnamefont{M.}~\bibnamefont{Koerner}},
  \bibinfo{author}{\bibfnamefont{A.}~\bibnamefont{Laeuchli}},
  \bibinfo{author}{\bibfnamefont{S.~R.} \bibnamefont{Manmana}},
  \bibinfo{author}{\bibfnamefont{I.~P.} \bibnamefont{MsCulloch}},
  \bibinfo{author}{\bibfnamefont{F.}~\bibnamefont{Michel}},
  \bibinfo{author}{\bibfnamefont{R.~M.} \bibnamefont{Noack}},
  \bibnamefont{et~al.}, \bibinfo{journal}{J. Phys. Soc. Jpn. Suppl.}
  \textbf{\bibinfo{volume}{74}}, \bibinfo{pages}{30} (\bibinfo{year}{2005}).

\bibitem[{\citenamefont{Albuquerque et~al.}(2007)\citenamefont{Albuquerque,
  Alet, Corboz, Dayal, Feiguin, Fuchs, Gamper, Gull, G{\"u}rtler, Honecker
  et~al.}}]{alps07}
\bibinfo{author}{\bibfnamefont{A.}~\bibnamefont{Albuquerque}},
  \bibinfo{author}{\bibfnamefont{F.}~\bibnamefont{Alet}},
  \bibinfo{author}{\bibfnamefont{P.}~\bibnamefont{Corboz}},
  \bibinfo{author}{\bibfnamefont{P.}~\bibnamefont{Dayal}},
  \bibinfo{author}{\bibfnamefont{A.}~\bibnamefont{Feiguin}},
  \bibinfo{author}{\bibfnamefont{S.}~\bibnamefont{Fuchs}},
  \bibinfo{author}{\bibfnamefont{L.}~\bibnamefont{Gamper}},
  \bibinfo{author}{\bibfnamefont{E.}~\bibnamefont{Gull}},
  \bibinfo{author}{\bibfnamefont{S.}~\bibnamefont{G{\"u}rtler}},
  \bibinfo{author}{\bibfnamefont{A.}~\bibnamefont{Honecker}},
  \bibnamefont{et~al.}, \bibinfo{journal}{J. Magn. Magn. Mater.}
  \textbf{\bibinfo{volume}{310}}, \bibinfo{pages}{1187} (\bibinfo{year}{2007}).

\bibitem[{\citenamefont{Krech et~al.}(1998)\citenamefont{Krech, Bunker, and
  Landau}}]{Krech98}
\bibinfo{author}{\bibfnamefont{M.}~\bibnamefont{Krech}},
  \bibinfo{author}{\bibfnamefont{A.}~\bibnamefont{Bunker}}, \bibnamefont{and}
  \bibinfo{author}{\bibfnamefont{D.~P.} \bibnamefont{Landau}},
  \bibinfo{journal}{Comput. Phys. Commun.} \textbf{\bibinfo{volume}{111}},
  \bibinfo{pages}{1} (\bibinfo{year}{1998}).

\bibitem[{\citenamefont{Landau et~al.}(2000)\citenamefont{Landau, Bunker,
  Evertz, Krech, and Tsai}}]{Landau00}
\bibinfo{author}{\bibfnamefont{D.~P.} \bibnamefont{Landau}},
  \bibinfo{author}{\bibfnamefont{A.}~\bibnamefont{Bunker}},
  \bibinfo{author}{\bibfnamefont{H.~G.} \bibnamefont{Evertz}},
  \bibinfo{author}{\bibfnamefont{M.}~\bibnamefont{Krech}}, \bibnamefont{and}
  \bibinfo{author}{\bibfnamefont{S.}~\bibnamefont{Tsai}},
  \bibinfo{journal}{Prog. Theor. Phys. Suppl.} \textbf{\bibinfo{volume}{138}},
  \bibinfo{pages}{423} (\bibinfo{year}{2000}).

\bibitem[{\citenamefont{Arrillaga and Watson}(2003)}]{FT_book}
\bibinfo{author}{\bibfnamefont{J.}~\bibnamefont{Arrillaga}} \bibnamefont{and}
  \bibinfo{author}{\bibfnamefont{N.~R.} \bibnamefont{Watson}},
  \emph{\bibinfo{title}{power system harmonics}} (\bibinfo{publisher}{John
  Wiley \& Sons Ltd.}, \bibinfo{year}{2003}).

\bibitem[{\citenamefont{Chen and Landau}(1994)}]{ChenLandau}
\bibinfo{author}{\bibfnamefont{K.}~\bibnamefont{Chen}} \bibnamefont{and}
  \bibinfo{author}{\bibfnamefont{D.~P.} \bibnamefont{Landau}},
  \bibinfo{journal}{Phys. Rev. B} \textbf{\bibinfo{volume}{49}},
  \bibinfo{pages}{3266} (\bibinfo{year}{1994}).

\bibitem[{\citenamefont{Evertz and Landau}(1996)}]{EvertzLandau}
\bibinfo{author}{\bibfnamefont{H.~G.} \bibnamefont{Evertz}} \bibnamefont{and}
  \bibinfo{author}{\bibfnamefont{D.~P.} \bibnamefont{Landau}},
  \bibinfo{journal}{Phys. Rev. B} \textbf{\bibinfo{volume}{54}},
  \bibinfo{pages}{12302} (\bibinfo{year}{1996}).

\bibitem[{\citenamefont{Wolff}(1989)}]{Wolff89}
\bibinfo{author}{\bibfnamefont{U.}~\bibnamefont{Wolff}},
  \bibinfo{journal}{Phys. Rev. Lett.} \textbf{\bibinfo{volume}{62}},
  \bibinfo{pages}{361} (\bibinfo{year}{1989}).

\bibitem[{\citenamefont{Suzuki}(1990)}]{Suzuki90}
\bibinfo{author}{\bibfnamefont{M.}~\bibnamefont{Suzuki}},
  \bibinfo{journal}{Phys. Lett. A} \textbf{\bibinfo{volume}{146}},
  \bibinfo{pages}{319} (\bibinfo{year}{1990}).

\bibitem[{\citenamefont{Suzuki}(1992)}]{Suzuki92}
\bibinfo{author}{\bibfnamefont{M.}~\bibnamefont{Suzuki}},
  \bibinfo{journal}{Phys. Lett. A} \textbf{\bibinfo{volume}{165}},
  \bibinfo{pages}{387} (\bibinfo{year}{1992}).

\bibitem[{\citenamefont{Yoshida}(1990)}]{Yoshida90}
\bibinfo{author}{\bibfnamefont{H.}~\bibnamefont{Yoshida}},
  \bibinfo{journal}{Phys. Lett. A} \textbf{\bibinfo{volume}{150}},
  \bibinfo{pages}{262} (\bibinfo{year}{1990}).

\end{thebibliography}

\end{document}